\documentclass[twocolumn]{aastex63}

\usepackage{float}
\usepackage{gensymb}
\usepackage{textcomp}
\usepackage[autopunct=true]{csquotes}
\usepackage{xcolor}

\shorttitle{GRB 190114C}
\shortauthors{Jordana-Mitjans et al.}

\begin{document}

\title{Lowly polarized light from a highly magnetized jet of GRB 190114C}

\correspondingauthor{N\'uria Jordana-Mitjans}
\email{N.Jordana@bath.ac.uk}

\author[0000-0002-5467-8277]{N. Jordana-Mitjans}
\affiliation{Department of Physics, University of Bath, \\
Claverton Down, Bath, BA2 7AY, UK}

\author[0000-0003-2809-8743]{C. G. Mundell}
\affiliation{Department of Physics, University of Bath, \\
Claverton Down, Bath, BA2 7AY, UK}

\author[0000-0001-7946-4200]{S. Kobayashi}
\affiliation{Astrophysics Research Institute, Liverpool John Moores University, \\
146 Brownlow Hill, Liverpool, L3 5RF, UK}

\author[0000-0003-3434-1922]{R. J. Smith}
\affiliation{Astrophysics Research Institute, Liverpool John Moores University, \\
146 Brownlow Hill, Liverpool, L3 5RF, UK}

\author[0000-0001-6869-0835]{C. Guidorzi}
\affiliation{Department of Physics and Earth Science, University of Ferrara, \\
via Saragat 1, I-44122, Ferrara, Italy}

\author[0000-0001-8397-5759]{I. A. Steele}
\affiliation{Astrophysics Research Institute, Liverpool John Moores University, \\
146 Brownlow Hill, Liverpool, L3 5RF, UK}

\author[0000-0002-4022-1874]{M. Shrestha}
\affiliation{Astrophysics Research Institute, Liverpool John Moores University, \\
146 Brownlow Hill, Liverpool, L3 5RF, UK}

\author[0000-0002-0908-914X]{A. Gomboc}
\affiliation{Center for Astrophysics and Cosmology, University of Nova Gorica, \\
Vipavska 13, 5000 Nova Gorica, Slovenia}

\author[0000-0002-5817-4009]{M. Marongiu}
\affiliation{Department of Physics and Earth Science, University of Ferrara, \\
via Saragat 1, I-44122, Ferrara, Italy}
\affiliation{ICRANet, Piazzale della Repubblica 10, I-65122, Pescara, Italy}

\author[0000-0002-0335-319X]{R. Martone}
\affiliation{Department of Physics and Earth Science, University of Ferrara, \\
via Saragat 1, I-44122, Ferrara, Italy}
\affiliation{ICRANet, Piazzale della Repubblica 10, I-65122, Pescara, Italy}

\author[0000-0003-3668-1314]{V. Lipunov}
\affiliation{Lomonosov Moscow State University, SAI, Physics Department, \\
13 Univeristetskij pr-t, Moscow 119991, Russia}

\author{E. S. Gorbovskoy}
\affiliation{Lomonosov Moscow State University, SAI, Physics Department, \\
13 Univeristetskij pr-t, Moscow 119991, Russia}

\author[0000-0002-7004-9956]{D. A. H. Buckley}
\affiliation{South African Astronomical Observatory \\
PO Box 9, Observatory 7935, Cape Town, South Africa}

\author[0000-0003-3767-7085]{R. Rebolo}
\affiliation{Instituto de Astrof\'isica de Canarias (IAC), \\
Calle V\'ia  L\'actea s/n, E-38200 La Laguna, Tenerife, Spain}

\author[0000-0002-2104-6687]{N. M. Budnev}
\affiliation{Applied Physics Institute, Irkutsk State University, \\
20, Gagarin blvd, 664003, Irkutsk, Russia}

\begin{abstract}

We report multi-color optical imaging and polarimetry observations of the afterglow of the first TeV- detected gamma-ray burst, GRB 190114C, using RINGO3 and MASTER II polarimeters. Observations begin 31 s after the onset of the GRB and continue until $\sim 7000\,$s post-burst. The light curves reveal a chromatic break at $\sim  400- 500\,$s --- with initial temporal decay $\alpha = 1.669 \pm 0.013$ flattening to $\alpha \sim 1$ post-break --- which we model as a combination of reverse and forward-shock components, with magnetization parameter $R_{\rm B} \sim 70$. The observed polarization degree decreases from $7.7 \pm 1.1\%$ to $2-4\%$ during $52-109\,$s post-burst and remains steady at this level for the subsequent $\sim 2000$-s, at constant position angle. Broadband spectral energy distribution modeling of the afterglow confirms GRB 190114C is highly obscured (A$_{\rm v, HG} = 1.49 \pm 0.12 \,$mag; N$_{\rm H, HG}= (9.0 \pm 0.3) \times 10^{22}\,$cm$^{-2}$). We interpret the measured afterglow polarization as intrinsically low and dominated by dust --- in contrast to  ${\rm P} >10\%$ measured previously for other GRB reverse shocks --- with a small contribution from polarized prompt photons in the first minute. We test whether 1st and higher-order inverse Compton scattering in a magnetized reverse shock can explain the low optical polarization and the sub-TeV emission but conclude neither is explained in the reverse shock Inverse Compton model. Instead, the unexpectedly low intrinsic polarization degree in GRB 190114C can be explained if large-scale jet magnetic fields are distorted on timescales prior to reverse shock emission.

\end{abstract}


\section{Introduction} \label{sec:intro}

Through the span of milliseconds to hundreds of seconds, gamma-ray bursts (GRBs) are the brightest sources of $\gamma$-ray photons in the universe. The accretion onto a compact object (e.g., a neutron star or a black hole) powers ultra-relativistic jets that via internal dissipation processes (e.g., internal shocks or magnetic reconnection) generate the characteristic and variable $\gamma$-ray prompt emission. Subsequently, the expanding ejecta collides with the circumburst medium producing a long-lived afterglow that can be detected at wavelengths across the electromagnetic spectrum \citep[e.g.,][]{1999PhR...314..575P,2002ARA&A..40..137M, 2004RvMP...76.1143P}.

GRB outflows provide a unique opportunity to probe the nature of GRB progenitors --- thought to involve the core-collapse of massive stars or the merger of compact stellar objects  \citep{1993ApJ...405..273W,2014ARA&A..52...43B,2017PhRvL.119p1101A,2017ApJ...848L..12A}  --- as well as acting as valuable laboratories for the study of relativistic jet physics (e.g. jet composition, energy dissipation, shock physics and radiation emission mechanisms) and their environments.

At the onset of the afterglow, two shocks develop: a forward shock that travels into the external medium and a short-lived reverse shock which propagates back into the jet \citep{1999ApJ...520..641S, 2000ApJ...545..807K}. The interaction between the outflow and the ambient medium can be quantified by the magnetization degree of the ejecta $\sigma_{\rm B}$, defined as the ratio of magnetic to kinetic energy flux. In a matter-dominated regime ($\sigma_{\rm B} \ll 1$; baryonic jet), the standard fireball model conditions are satisfied and internal shocks are plasma-dominated \citep{1994ApJ...430L..93R}. For increasing $\sigma_{\rm B}$, the reverse shock becomes stronger until it reaches a maximum at $\sigma_{\rm B} \sim 0.1$ and it becomes progressively weaker and likely suppressed for $\sigma_{\rm B} \gtrsim 1$ \citep{2003ApJ...595..950Z, 2004A&A...424..477F, 2005ApJ...628..315Z, 2008A&A...478..747G}. For an outflow highly magnetized at the deceleration radius ($\sigma_{\rm B} \gg 1$; Poynting-flux jet), the magnetic fields are dynamically dominant, prompt emission is understood in terms of magnetic dissipation processes and the ejecta carries globally ordered magnetic fields \citep{1994MNRAS.267.1035U, 2001AandA...369..694S, 2003astro.ph.12347L}.

Observations of the optical afterglow show low or no polarization at late times ($\sim$1 day) when the forward shock --- powered by shocked ambient medium --- dominates the light curve (e.g., \citealt{1999A&A...348L...1C}). In contrast, the prompt and early-time afterglow emission from the reverse shock are sensitive to the properties of the central engine ejecta. At this stage, different polarization signatures are predicted for magnetic and baryonic jet models. In a Poynting-flux dominated jet, the early-time emission is expected to be highly polarized due to the presence of primordial magnetic fields advected from the central engine \citep{2003ApJ...594L..83G, 2003ApJ...597..998L,2004A&A...424..477F, 2005ApJ...628..315Z}. In a baryonic jet, tangled magnetic fields locally generated in shocks are randomly oriented in space giving rise to unpolarized emission for on-axis jets \citep{1999ApJ...526..697M} or mild polarization detections for edge-on jets \citep{1999MNRAS.309L...7G,1999ApJ...524L..43S}. Therefore, early-time polarization measurements of the afterglow are crucial for diagnosing its composition and discriminating between competing jet models.

Polarization measurements are technically challenging and reverse shock detections remain rare (e.g., \citealt{2014ApJ...785...84J}). However, the advent of autonomous optical robotic telescopes and real-time arcminute localization of GRBs has made these observations feasible \citep{2005SSRv..120..143B, 2004SPIE.5489..679S}.

The first early-time polarization measurement in the optical was achieved with GRB 060418 \citep{2007Sci...315.1822M}. The fast response of the polarimeter allowed observations during the deceleration of the blast wave, beginning $203\,$s after the GRB. The upper limit of $8\%$ at this time favored either reverse shock suppression due to a highly magnetized ejecta or ruled out the presence of large-scale ordered magnetic fields with dominant reverse shock emission.

The measurement of $10 \pm 1 \%$ during the steep decay of GRB 090102 reverse shock --- measured only $160\,$s post-burst --- was the first evidence that  large-scale ordered magnetic fields are present in the fireball \citep{2009Natur.462..767S}. The $6^{+3} _{-2} \%$ and $6^{+4} _{-3} \%$ detection during the rise and decay of GRB 101112A afterglow and the $13 ^{+13} _{-9} \%$ measurement during the rapid rise of GRB 110205A afterglow indicated reverse shock contribution \citep{2011ApJ...743..154C, 2017ApJ...843..143S}. GRB 120308A polarization gradual decrease from $28 \pm 4\%$ to $16 ^{+5} _{-4} \%$ revealed that these large-scale fields could survive long after the deceleration of the fireball \citep{2013Natur.504..119M}. The time-sampled polarimetry for both GRB 101112A and GRB 120308A indicated that the polarization position angle remained constant or rotated only gradually, consistent with stable, globally ordered magnetic fields in a relativistic jet. The first detection of polarized prompt optical emission was reported by \cite{2017Natur.547..425T} for GRB 160625B. 

In combination, the existence of bright reverse shock emission theoretically requires a mildly magnetized jet and the early-time polarization studies favor the presence of primordial magnetic fields advected from the central engine.

GRB 190114C is the first of its kind to be detected by the Major Atmospheric Gamma Imaging Cherenkov Telescope (MAGIC) at sub-TeV energies \citep{2019ATel12390....1M}, challenging GRB models for the production of GeV-TeV energies \citep{2019A&A...626A..12R,2019ApJ...883..162F, 2019ApJ...880L..27D,2019ApJ...884..117W,2019arXiv190910605A}. Moreover, GRB 190114C prompt emission was followed by a very bright afterglow, which makes it an interesting candidate for time-resolved polarimetric observations at early times \citep{2013Natur.504..119M, 2017Natur.547..425T,2017ApJ...843..143S}. 

\begin{figure*}[ht!]
\plotone{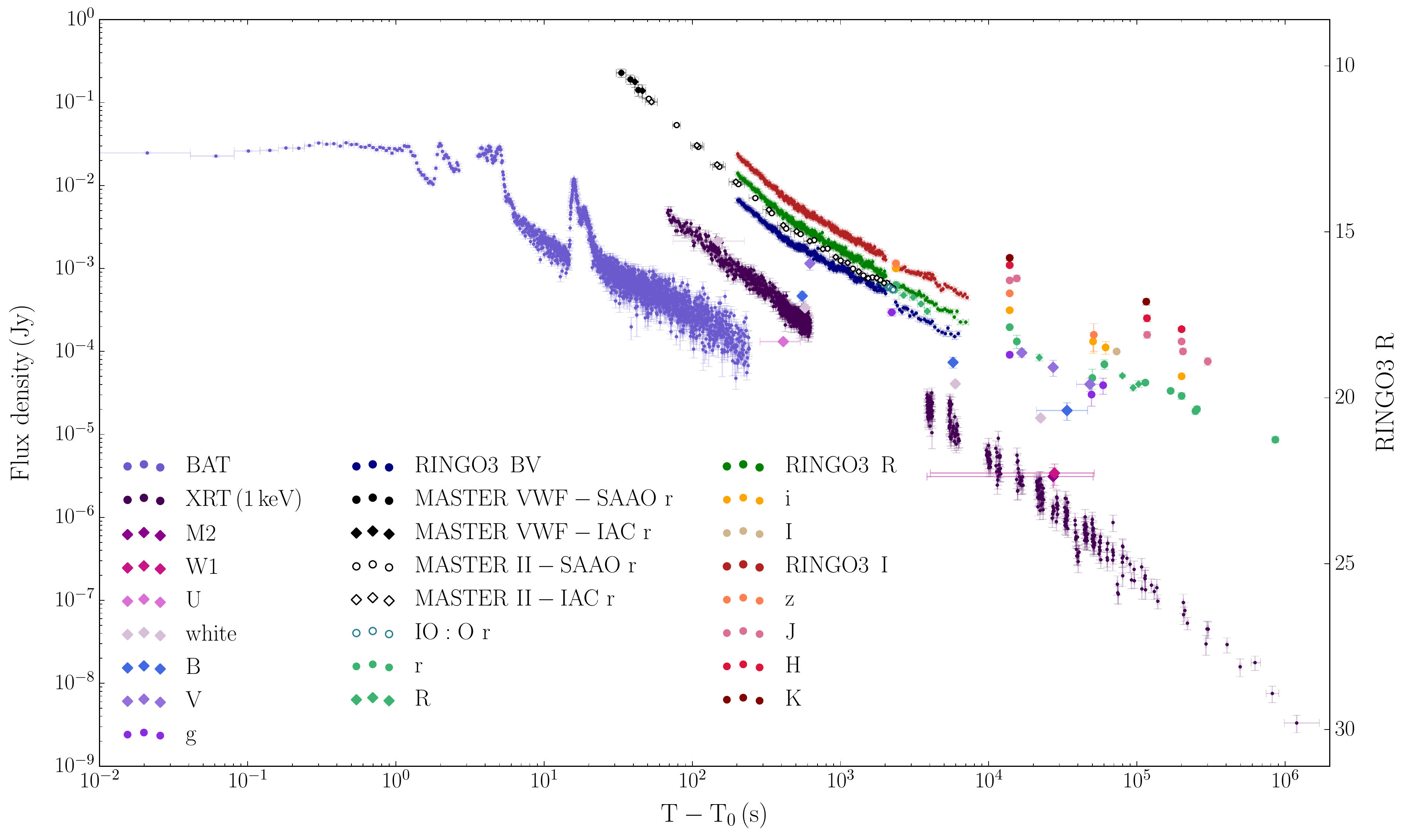}
\caption{GRB 190114C multi-wavelength light curves with Swift BAT, Swift XRT, MASTER-SAAO/IAC VWF r-equivalent, MASTER-SAAO/IAC II r-equivalent, LT RINGO3 BV/R/I and LT IO:O r bands. Swift data is obtained from the web interface provided by Leicester University \citep{2009MNRAS.397.1177E}: BAT data is binned to signal-to-noise 5 and the absorbed 0.3-10 keV XRT light curve is converted to flux density at 1 keV. For completeness, we include the UV/optical/infrared observations reported in GCNs from UVOT \citep{2019GCN.23725....1S}, NOT \citep{2019GCN.23695....1S}, OASDG \citep{2019GCN.23699....1I}, GROND \citep{2019GCN.23702....1B}, REM \citep{2019GCN.23754....1D}, McDonald observatory \citep{2019GCN.23717....1I}, LSGT \citep{2019GCN.23732....1K}, GRowth-India \citep{2019GCN.23733....1K}, KMTNet \citep{2019GCN.23734....1K}, UKIRT \citep{2019GCN.23757....1I}, CHILESCOPE \citep{2019GCN.23741....1M, 2019GCN.23746....1M, 2019GCN.23787....1M}, RTT150 \citep{2019GCN.23766....1B}, ePESSTO NTT \citep{2019GCN.23748....1R}, RATIR \citep{2019GCN.23751....1W} and HCT \citep{2019GCN.23742....1K,2019GCN.23798....1S}. GCNs observations do not include filter corrections. In the x-axis, T$_0$ corresponds to BAT trigger time; in the y-axis, the flux density is converted to RINGO3 R magnitude.}
\label{fig:LC_GRB190114C_all}
\end{figure*}

In this work, we present the early-time multicolor optical imaging polarimetric observations of GRB 190114C with the RINGO3 three-band imaging polarimeter \citep{2012SPIE.8446E..2JA} mounted on the 2-m autonomous robotic optical Liverpool Telescope \citep[LT;][]{2004SPIE.5489..679S, 2006PASP..118..288G} and with the fully robotic 0.4-m MASTER-SAAO/IAC II telescopes from the MASTER Global Robotic Net \citep{2010AdAst2010E..30L,2012ExA....33..173K}. The paper is structured as follows: the data reduction of Liverpool Telescope and MASTER observations are reported in Section \ref{sec:observations}; in Section \ref{sec:Results}, we characterize the temporal, polarimetric and spectral properties of the burst in three optical bands with observations starting $201\,$s post-burst and in a white band since $30.7\,$s; in Section \ref{sec:th_model}, we model the optical afterglow with a reverse-forward shock model; in Section \ref{sec:discussion}, we discuss reverse shock Synchrotron-Self-Compton as a possible mechanism for the sub-TeV detection and we infer the strength and structure of the magnetic field in the outflow. The results are summarized in Section \ref{sec:conclusions}. Throughout this work, we assume flat $\Lambda$CDM cosmology $\Omega_{\rm m} = 0.32$, $\Omega_{\Lambda} = 0.68$ and $h=0.67$, as reported by \cite{2018arXiv180706209P}. We adopt the convention $F_{\nu} \propto t ^{-\alpha} \nu ^{-\beta} $, where $\alpha$ is the temporal index and $\beta$ is the spectral index. Uncertainties are quoted at $1\sigma$ confidence level unless stated otherwise.

\section{Observations and Data Reduction} \label{sec:observations}

On 2019 January 14 at T$_0=$20:57:03$ \,$UT, Swift Burst Alert Telescope \citep[BAT;][]{2005SSRv..120..143B} triggered an alert for the GRB candidate 190114C and immediately slewed towards its position \citep{2019GCN.23688....1G}. Other telescopes also reported the detection of GRB 190114C $\gamma$-ray prompt as a multi-peaked structure: Konus-Wind \citep[KW;][]{2019GCN.23737....1F}, Fermi Gamma-ray Burst Monitor \citep[GBM;][]{2019GCN.23707....1H}, Fermi Large Area Telescope \citep[LAT;][]{2019GCN.23709....1K}, Astro-Rivelatore Gamma a Immagini Leggero \citep[AGILE;][]{2019GCN.23712....1U}, INTErnational Gamma-Ray Astrophysics Laboratory \citep[INTEGRAL;][]{2019GCN.23714....1M} and the Hard X-ray Modulation Telescope \citep[Insight-HXMT/HE;][]{2019GCN.23716....1X}. At T$_0 + 50\,$s, the Cherenkov telescope MAGIC detected the burst at energies higher than $300$ GeV with a significance of $>20\sigma$ \citep{2019ATel12390....1M}. 

Due to the different spectral coverage of the detectors and the presence of soft extended emission \citep{2019GCN.23707....1H, 2019GCN.23714....1M, 2019GCN.23737....1F}, the long $\gamma$-ray prompt was observed to last T$_{90} = 362 \pm 12\,$s in the 15-350 keV band \citep[BAT;][]{2019GCN.23724....1K}, T$_{90} = 116\,$s in the 50-300 keV band \citep[GBM;][]{2019GCN.23707....1H}, T$_{90} = 15.7\,$s in the 200-3000 keV band \citep[Insight-HXMT/HE;][]{2019GCN.23716....1X} and T$_{90} = 6.2\,$s in the 0.4-100 MeV band \citep[AGILE;][]{2019GCN.23712....1U}. KW analysis reported an energy peak E$_{\rm peak} =  646 \pm 16 \, $keV, an isotropic energy E$_{\rm iso} = (2.40 \pm 0.05) \times 10^{53}\,$erg, a peak luminosity L$_{\rm iso}= (1.67 \pm 0.05) \times 10^{53}\,$erg/s and pointed out that these values follow the Amati-Yonetoku relation within $1\sigma$ \citep{2019GCN.23737....1F}.

Seconds to days after the burst, GRB 190114C afterglow was observed at wavelengths from the X-rays to the infrared (see Figure \ref{fig:LC_GRB190114C_all}; references therein) and down to radio frequencies \citep{2019GCN.23728....1L, 2019ApJ...878L..26L, 2019GCN.23726....1A, 2019GCN.23745....1S, 2019GCN.23750....1V, 2019GCN.23760....1T, 2019GCN.23762....1C}. The fastest response to BAT trigger was from the MASTER-SAAO VWF camera at T$_0 + 30.7\,$s with a $\sim 10.51 \pm 0.12\,$mag detection in the optical (see Section \ref{sec:MASTER_follow}). Later observations started at T$_0+ 67\,$s, T$_0 + 74\,$s and T$_0 + 201\,$s with the Swift X-ray Telescope \citep[XRT;][]{2019GCN.23706....1D}, the 0.3-m Ultraviolet/Optical Telescope \citep[UVOT;][]{2019GCN.23725....1S} and the 2-m Liverpool Telescope (see Section \ref{sec:LT_follow}), respectively. A spectroscopic redshift of $0.4245\pm 0.0005$ was measured by the 10.4-m GTC telescope and confirmed by the 2.5-m Nordic Optical Telescope \citep{2019GCN.23695....1S, 2019GCN.23708....1C}. Additionally, a supernova component was detected $15\,$days after the burst, confirming a collapsar origin for GRB 190114C \citep{2019GCN.23983....1M}.

\subsection{Follow-up Observations by the Liverpool Telescope} \label{sec:LT_follow}

The 2-m robotic Liverpool Telescope \citep[LT;][]{2004SPIE.5489..679S, 2006PASP..118..288G} started observing the field $201\,$s after the burst with the multi-wavelength imager and polarimeter RINGO3. For a typical GRB follow-up, the telescope autonomously schedules a series of $3 \times 10$-min observations with RINGO3 followed by a $6 \times 10\,$-s sequence with the r-SDSS filter of the Optical Wide Field Camera\footnote{https://telescope.livjm.ac.uk/TelInst/Inst/IOO/} (IO:O). Due to GRB 190114C exceptional brightness, an additional $8 \times 10$-min integrations were triggered with RINGO3 after IO:O observations.

RINGO3 is a fast-readout optical polarimeter that simultaneously provides polarimetry and imaging in three optical/infrared bands \citep{2012SPIE.8446E..2JA}. The instrument design includes a rotating polaroid that continuously images a $4 \times 4 \,$arcmin field at 8 rotor positions. Each RINGO3 $10\,$-min primary data product is composed of $10 \times 1$-min exposure frames. These frames are automatically generated by the LT reduction pipeline\footnote{https://telescope.livjm.ac.uk/TelInst/Pipelines/} which co-adds the individual $2.34\,$-s frames that correspond to a single polaroid rotation and corrects for bias, darks, and flats. For photometry, we integrate the counts over all polaroid positions (see Section \ref{sec:phot_reduc}); for polarimetry, we analyze the relative intensity of the source at the 8 angle positions of the polaroid (see Section \ref{sec:pola_reduc}).

\subsubsection{Frame Binning and Three-Band Light Curve Extraction}\label{sec:phot_reduc}

We use aperture photometry to compute the source flux; in particular, we employ the {\sc Astropy Photutils} package \citep{2016ascl.soft09011B}. The brightness of the OT during RINGO3 observations provided high signal-to-noise ratio even at high-temporal resolution; the source was detected at a signal-to-noise of $\gtrsim 60$ in each of the first $\sim 10 \times 2.34\,$-s frames. Due to the fading nature of the afterglow, the signal-to-noise of the detection rapidly drops for the following observations (e.g., $200\,$s later, the signal-to-noise of each $2.34\,$-s frame had decreased to $\sim 30$).  By $\sim {\rm T}_0 + 2000\,$s, the source was detected in the 1-min frames at signal-to-noise $\sim 25$. Consequently, our data choice is to use the $2.34\,$-s RINGO3 frames for the first $30\,$-min of observations to allow high-temporal resolution and then, the $1\,$-min exposures for the succeeding 1.3 hours.

At later times, when the OT has faded, we dynamically co-add frames and accept measurements with a $\geq 20$ signal-to-noise detection. With this signal-to-noise criteria, $\gtrsim {\rm T}_0 + 700\,$s measurements are the result of co-adding frames. Integrating at different signal-to-noise ratios does not change the light curve general features: $\ll 20$ signal-to-noise integrations show additional internal structure that is statistically not significant at $3\sigma$ level; $\gg 20$ signal-to-noise ratios further smooth minor features and reject fainter OT detections at later times.

To test for instrument stability during RINGO3 observations, we study the flux variability of the only star in the field (CD-27 1309; $\sim 11\,$mag star). Using the OT binning, CD-27 1309 photometry presents a $ \sim 0.01\,$mag deviation from the mean in all bands (or $\sim 1\%$ in flux).

The Optical Wide Field Camera (IO:O) observations started $34.7\,$min post-burst with the r filter. Given that the OT signal-to-noise ratio is $\sim 40$ for each of the $10\,$-s frames, we derive its flux from the 6 exposures, individually. IO:O r magnitudes are standardized using five $\sim 14-15 \,$mag stars from Pan-STARRS DR1 catalogue \citep{2016arXiv161205560C}. In Table \ref{tab:phot} and Figure \ref{fig:LC_GRB190114C_all}, we present the IO:O r filter photometry. The IO:O light curve is corrected for the mean Galactic extinction A$_{\rm r} = 0.034 \pm 0.001\,$mag \citep[E$_{\rm B-V, MW} =0.0124 \pm 0.0005$ is derived\footnote{https://irsa.ipac.caltech.edu/applications/DUST/} from a $5 \times 5\,$arcmin field statistic;][]{1998ApJ...500..525S} but not for host galaxy extinction (see Section \ref{sec:SEDs_broadband}).

\subsubsection{RINGO3 Bandpasses Standardization} \label{sec:RINGO3bandpasses}

\begin{figure}[ht!]
\plotone{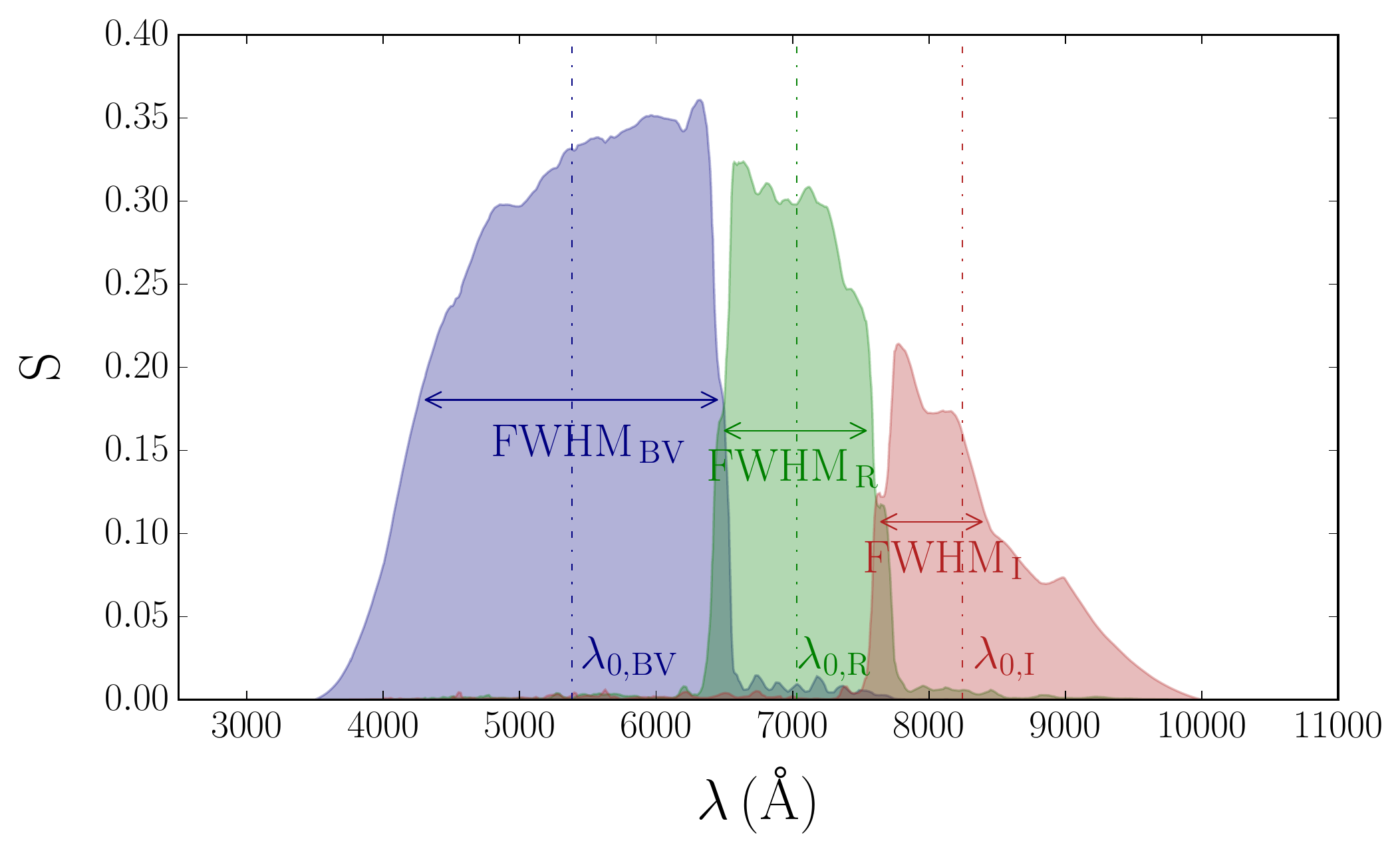}
\caption{Photonic response functions of RINGO3 BV/R/I bandpasses, which encompass the total instrument throughput (including atmospheric extinction).}
\label{fig:RINGO3_bandpasses}
\end{figure}

After RINGO3 polaroid, the light is split by two dichroic mirrors in three beams that are simultaneously recorded by three EMCCD cameras \citep{2012SPIE.8446E..2JA}. In Figure \ref{fig:RINGO3_bandpasses}, we derive the photonic response function of RINGO3 instrument which accounts for atmospheric extinction \citep{PalmaTech}, telescope optics\footnote{https://telescope.livjm.ac.uk/Pubs/LTTechNote1\_Telescope Throughput.pdf}, instrument dichroics\footnote{https://telescope.livjm.ac.uk/TelInst/Inst/RINGO3/}, lenses \citep{ArnoldThesis}, filters\footnote{https://www.meadowlark.com/versalight-trade-polarizer-p-79?mid=6\#.Wun27maZMxE}\footnote{https://www.thorlabs.com/newgrouppage9.cfm?objectgroup\_ id=870} transmission and the quantum efficiency of the EMCCDs \citep{ArnoldThesis}. The total throughput results in three broad bandpasses with the following mean photonic wavelengths $\lambda_{\rm 0, \, \lbrace BV, R, I \rbrace} = 5385\,{\rm \AA}, 7030\,{\rm \AA}, 8245\,{\rm \AA} \,$ and full-widths-at-half-maximum FWHM$_{\rm \lbrace BV, R, I \rbrace} = 2232\,{\rm \AA}, 1130\,{\rm \AA}, 835\,{\rm \AA}$.

Because of the different spectral coverage of RINGO3 bandpasses relative to other photometric systems and the $\sim 0.02-0.05\,$mag photometric precision, we standardize RINGO3 magnitudes in Vega system following \cite{1953ApJ...117..313J} procedure. Observations of four unreddened A0 type stars (HD 24083, HD 27166, HD 50188, HD 92573) and the GRB 190114C field were submitted via LT phase2UI\footnote{https://telescope.livjm.ac.uk/PropInst/Phase2/} using the same instrumental set-up of the night of the burst and autonomously dispatched on the nights of 2019 January 30-31. We standardize the magnitudes in RINGO3 system using CD-27 1309 star, which adds $\sim 0.05\,$mag uncertainty to the photometry.

Taking into account the notation $m = -2.5 \log(F_{\nu}) + C_{\nu}$ with $F_{\nu}$ in erg cm$^{-2}$ s$^{-1}$ Hz$^{-1}$ \citep[e.g.,][]{1998yCat..33330231B, 2012PASP..124..140B}, we compute the magnitude to flux density conversion $C_{\nu}$ by deriving the mean flux density $F_{\nu}$ of Vega star ($\alpha$ Lyr) composite spectrum\footnote{We use alpha\_lyr\_stis\_008.fits spectrum version from CALSPEC archive} through each RINGO3 band \citep{2014PASP..126..711B}. We set $m = 0$ for all bandpasses and we obtain $C_{\rm \nu,  \lbrace BV, R, I \rbrace} = -48.60, -48.90, -49.05$.

\begin{deluxetable}{ccccccccc}[h!]
\tablecaption{GRB 190114C LT observations with RINGO3 BV/R/I, IO:O r bands and MASTER VWF/MASTER II observations in a r-equivalent band. \label{tab:phot}}
\tablecolumns{8}
\tablewidth{0pt}
\tablehead{
\colhead{Band} & \colhead{t$_{\rm mid}$} & \colhead{t$_{\rm exp}/2$}  & \colhead{mag} & \colhead{mag$_{\rm \, err}$}  & \colhead{F$_{\nu}$} & \colhead{F$_{\nu \, \rm{err}}$} \\
\colhead{} & \colhead{(s)} & \colhead{(s)} & \colhead{} & \colhead{} & \colhead{(Jy)} & \colhead{(Jy)} }
\startdata
BV &    202.5 &      1.2 &    14.33 &     0.06 & 6.64e-03 &  3.8e-04\\
BV &    204.8 &      1.2 &    14.36 &     0.06 & 6.49e-03 &  3.7e-04\\
BV &    207.2 &      1.2 &    14.32 &     0.06 & 6.70e-03 &  3.9e-04\\
BV &    209.5 &      1.2 &    14.36 &     0.06 & 6.49e-03 &  3.7e-04\\
BV &    211.9 &      1.2 &    14.38 &     0.06 & 6.37e-03 &  3.7e-04\\
BV &  ... & ... & ... & ... & ... & ...
\enddata
\tablecomments{t$_{\rm mid}$ corresponds to the mean observing time and t$_{\rm exp}$ to the length of the observation window. Magnitudes and flux density values are corrected for atmospheric and Galactic extinction. Table \ref{tab:phot} is published in its entirety in machine-readable format. A portion is shown here for guidance regarding its form and content.}
\end{deluxetable}

\begin{deluxetable*}{cccccccccccc}[t!]
\tablecaption{GRB 190114C polarization observations with LT RINGO3 BV/R/I bands and MASTER II white band.
\label{tab:pol}}
\tablecolumns{12}
\tablewidth{0pt}
\tablehead{
\colhead{Band} & \colhead{t$_{\rm mid}$} & \colhead{t$_{\rm exp}/2$} &  \colhead{SNR} &\colhead{q} &\colhead{q$_{\rm err}$} & \colhead{u} & \colhead{u$_{\rm err}$} & \colhead{P}  & \colhead{P$_{\rm err}$} & \colhead{$\theta$} & \colhead{$\theta_{\rm err}$} \\
\colhead{} & \colhead{(s)} & \colhead{(s)} & \colhead{} & \colhead{} & \colhead{} & \colhead{} & \colhead{} & \colhead{($\%$)}  & \colhead{($\%$)} & \colhead{(\textdegree)} & \colhead{(\textdegree)} }
\startdata
BV & 223.5 &  22.1 &  71 & -0.020 & 0.022 & 0.018 & 0.011 &   2.7 & $_{-  1.4} ^{+  1.7}$ &    69 & $_{-   18} ^{+   18}$ \\
BV & 283.3 &  39.7 &  70 & -0.019 & 0.022 & 0.009 & 0.011 &   2.1 & $_{-  1.3} ^{+  1.7}$ &    77 & $_{-   25} ^{+   25}$ \\
BV & 433.4 & 112.4 &  70 & -0.020 & 0.022 & 0.019 & 0.011 &   2.8 & $_{-  1.4} ^{+  1.7}$ &    68 & $_{-   17} ^{+   17}$ \\
BV & 671.5 & 127.7 &  50 & -0.027 & 0.031 & 0.023 & 0.016 &   3.6 & $_{-  2.0} ^{+  2.4}$ &    70 & $_{-   19} ^{+   19}$ \\
BV & 1117.2 & 298.9 &  54 & -0.022 & 0.029 & 0.027 & 0.014 &   3.5 & $_{-  1.8} ^{+  2.1}$ &    64 & $_{-   18} ^{+   18}$ \\
BV & 1734.1 & 298.9 &  38 & -0.027 & 0.041 & 0.036 & 0.020 &   4.5 & $_{-  2.5} ^{+  3.0}$ &    63 & $_{-   20} ^{+   20}$ \\
\hline
R & 215.3 &  13.9 &  70 & -0.025 & 0.022 & 0.029 & 0.011 &   3.8 & $_{-  1.5} ^{+  1.7}$ &    65 & $_{-   12} ^{+   12}$ \\
R & 245.8 &  18.6 &  71 & -0.029 & 0.022 & 0.010 & 0.011 &   3.0 & $_{-  1.4} ^{+  1.6}$ &    80 & $_{-   16} ^{+   16}$ \\
R & 293.8 &  31.5 &  70 & -0.028 & 0.022 & 0.023 & 0.011 &   3.6 & $_{-  1.5} ^{+  1.6}$ &    70 & $_{-   13} ^{+   13}$ \\
R & 386.5 &  63.2 &  70 & -0.019 & 0.022 & 0.014 & 0.011 &   2.4 & $_{-  1.4} ^{+  1.7}$ &    71 & $_{-   21} ^{+   21}$ \\
R & 623.4 & 175.8 &  61 & -0.024 & 0.026 & 0.007 & 0.013 &   2.5 & $_{-  1.5} ^{+  1.9}$ &    82 & $_{-   24} ^{+   23}$ \\
R & 1117.2 & 298.9 &  45 & -0.029 & 0.035 & 0.008 & 0.017 &   3.0 & $_{-  2.0} ^{+  2.6}$ &    83 & $_{-   27} ^{+   27}$ \\
R & 1734.1 & 298.9 &  31 & -0.006 & 0.051 & 0.020 & 0.026 &   2.1 & $_{-  1.6} ^{+  4.0}$ &    53 & $_{-   43} ^{+  110}$ \\
\hline
I & 215.2 &  13.9 &  70 & -0.036 & 0.022 & 0.023 & 0.011 &   4.2 & $_{-  1.5} ^{+  1.6}$ &    74 & $_{-   11} ^{+   11}$ \\
I & 245.7 &  18.6 &  70 & -0.018 & 0.022 & 0.003 & 0.011 &   1.8 & $_{-  1.2} ^{+  1.7}$ &    86 & $_{-   29} ^{+   29}$ \\
I & 292.6 &  30.3 &  70 & -0.022 & 0.022 & 0.020 & 0.011 &   3.0 & $_{-  1.4} ^{+  1.7}$ &    69 & $_{-   16} ^{+   16}$ \\
I & 380.6 &  59.7 &  70 & -0.012 & 0.022 & 0.012 & 0.011 &   1.7 & $_{-  1.2} ^{+  1.7}$ &    67 & $_{-   32} ^{+   34}$ \\
I & 618.7 & 180.5 &  63 & -0.024 & 0.025 & 0.007 & 0.012 &   2.5 & $_{-  1.5} ^{+  1.9}$ &    82 & $_{-   22} ^{+   22}$ \\
I & 1117.2 & 298.9 &  45 & -0.007 & 0.035 & 0.003 & 0.017 &   0.8 & $_{-  0.5} ^{+  2.9}$ &    78 & $_{-   69} ^{+   92}$ \\
I & 1734.1 & 298.9 &  33 & 0.019 & 0.048 & -0.025 & 0.024 &   3.2 & $_{-  2.3} ^{+  3.7}$ &   154 & $_{-  148} ^{+   23}$ \\
\hline
White & 52.0 & 6.1 & 264 & -0.076 & 0.005 & -0.015 & 0.005 & 7.7 & $^{+1.1} _{-1.1}$ & 96$^{a}$ & $^{+4} _{-4}$ \\
White & 78.4  & 5.0  & 147 & -  & - &  - & -  &  $>2.2$ & $_{-  0.6} ^{+  0.6}$ & - & - \\
White & 108.6 & 8.8 & 135 & -0.020 & 0.012 & 0.003 & 0.012 & 2.0 & $^{+2.6} _{-1.5}$ & 85$^{a}$ & $^{+44} _{-43}$ \\
White & 149.6 & 12.3 & 103 & 0.012 & 0.014 & 0.002 & 0.014 & 1.2 & $^{+3.1} _{-0.8}$ & 4$^{a}$ & $^{+175} _{-2}$ \\
White & 200.7 & 16.0 & 78 & 0.021 &
0.019 & 0.003 & 0.019 & 2.1 & $^{+4.3} _{-1.5}$ & 4$^{a}$ & $^{+174} _{-3}$
\enddata
\tablecomments{t$_{\rm mid}$ corresponds to the mean observing time, t$_{\rm exp}$ to the length of the observation window and SNR to the signal-to-noise ratio of the OT. The Stokes parameters q-u, the polarization degree P and the polarization angle $\theta$ are corrected for instrumental effects. P and $\theta$ uncertainties are quoted at $2\sigma$ confidence level.}
\tablenotetext{a}{$\theta$ is not calibrated with polarimetric standard stars.}
\end{deluxetable*}

In Table \ref{tab:phot} and Figure \ref{fig:LC_GRB190114C_all}, we present the GRB 190114C absolute flux calibrated photometry of RINGO3 BV/R/I bands. All three light curves start at a mean time ${\rm T}_0 + 202.5\,$s. R and I band photometry ends at $\sim 7000\,$s post-burst. For the BV band, the stacking does not reach the signal-to-noise $\geq 20$ threshold for the last $\sim 800\,$-s of observations and therefore, the photometry is discarded. Magnitudes and flux density are corrected for atmospheric extinction with $M_{\rm c, \, \lbrace BV, R, I \rbrace}= 0.14\,$mag, $0.04\,$mag, $0.02\,$mag and $F_{\rm c, \, \lbrace BV, R, I \rbrace} = 0.89, 0.96, 0.98$, respectively, which we compute from a weighted mean of the bandpasses throughput and the theoretical atmospheric extinction of the site \citep{PalmaTech}. We also correct for the mean Galactic extinction, A$_{\rm  \lbrace BV, R, I \rbrace}/$E$_{\rm B-V}= 3.12, 2.19, 1.73$ with E$_{\rm B-V, MW} =0.0124 \pm 0.0005$ \citep{1998ApJ...500..525S}, which we derive using \cite{1992ApJ...395..130P} Milky Way dust extinction profile. The light curves are not corrected for host galaxy extinction (see Section \ref{sec:SEDs_broadband}).

\subsubsection{RINGO3 Polarization Calibration}  \label{sec:pola_reduc}

To derive the polarization of a source with RINGO3 instrumental configuration, we first compute the OT flux at each rotor position of the polaroid with aperture photometry using {\sc Astropy Photutils} package \citep{2016ascl.soft09011B}. The flux values are converted to Stokes parameters q-u following \cite{2002A&A...383..360C} procedure and then to polarization degree and angle. Following \cite{2016MNRAS.458..759S} methodology to correct for RINGO3 polarization instrumental effects, we use 44 observations of BD +32 3739, BD +28 4211, HD 212311 unpolarized stars and 41 observations of HILT 960, BD +64 106 polarized stars for each band. Due to the positive nature of polarization\footnote{The polarization degree and angle are related to the Stokes parameters as $p = \sqrt{q^2 + u^2}$ and $\theta = 0.5 \arctan(u/q)$}, measurements are not normally distributed in the low signal-to-noise and low polarization regime \citep{1985A&A...142..100S}. Consequently, to derive the confidence levels in the Stokes parameters and polarization, we perform a Monte Carlo error propagation starting with 10$^{5}$ simulated flux values for each rotor position.

Following \cite{2013Natur.504..119M}, we initially infer the polarization of the source with a single measurement, with maximum signal-to-noise. By co-adding the $2.34\,$-s frames of the first $10\,$-min epoch, we obtain a signal-to-noise detection of $\sim 130$ corresponding to a mean time of $\sim 321 \pm 120\,$s. From this estimate, we derive a polarization degree at $2\sigma$ confidence level P$_{\rm \lbrace BV, R, I \rbrace}= 2.2_{-  0.8} ^{+  0.9} \%$, $2.9_{-  0.8} ^{+  0.9} \%$, $2.4_{-  0.8} ^{+  0.9} \%$, angle $\theta_{\rm \lbrace BV, R, I \rbrace}= 81 \pm 12$\textdegree, $70 \pm 9$\textdegree, $71 \pm 11$\textdegree and Stokes parameters q$_{\rm \lbrace BV, R, I \rbrace}= -0.021 \pm 0.006$, $-0.022 \pm 0.006$, $-0.019 \pm 0.006$, u$_{\rm \lbrace BV, R, I \rbrace} = 0.007 \pm 0.006$, $0.019 \pm 0.006$, $0.015 \pm 0.006$. In this paper, we quote $2\sigma$ confidence levels for the polarization degree P and angle $\theta$ because it better reflects the non-gaussian behavior of polarization in the low degree regime.

Polarization is a vector quantity, variation in either or both degree/angle on timescales shorter than $\Delta t \sim 240 \,$s can result in a polarization detection of lower degree. To check for variability in polarization on timescales $\Delta t < 240 \,$s, we dynamically co-add the $2.34\,$-s frames at a lower signal-to-noise such that they reach a threshold of $\sim 70$. With this choice, we can claim polarization variability at $3\sigma$ confidence level if we measure a change in the polarization degree of $\gtrsim 3\%$. Integrations at higher and lower signal-to-noise ratios reproduce the results within $1\sigma$; however, because we estimate polarization to be $\sim 2-3\%$, $\ll 50$ signal-to-noise integrations are dominated by instrumental noise and are essentially upper limits. The remaining frames of the first $10\,$-min epoch and the following $2 \times 10\,$-min are co-added as individual measurements to ensure a maximal signal-to-noise. We do not use the next $8 \times 10 \,$-min epochs because the signal-to-noise declines below $\sim 10$ and falls within the instrument sensitivity; the instrumental noise is dominating polarization detections of $\lesssim 6\%$.

In Table \ref{tab:pol}, we present the Stokes parameters and the polarization degree and angle for the three RINGO3 bandpasses. To check for instrument stability, we calculate the star CD-27 1309 polarization using the OT binning choice. CD-27 1309 manifests deviations of $\sim 0.15 \%$ from the mean. Due to the sensitivity of polarization with the photometric aperture employed, we check that apertures within $1.5-3$FWHM yield polarization measurements compatible within $1\sigma$ for both CD-27 1309 and the OT.

\subsection{Follow-up Observations by the MASTER Global Robotic Net} \label{sec:MASTER_follow}

The earliest detection of GRB 190114C afterglow was done $30.7\,$s post-burst with the Very Wide-Field (VWF) camera from MASTER-SAAO observatory, which is part of the MASTER Global Robotic Net \citep{2010AdAst2010E..30L,2012ExA....33..173K}. About $8\,$s later, MASTER-IAC VWF also detected the OT. The VWF camera enables wide-field coverage in a white band and constant sky imaging every $5\,$s, which is crucial for GRB prompt detections \citep{2010AdAst2010E..62G}. 

At $\sim 47\,$s post-burst, MASTER-SAAO and MASTER-IAC observatories started nearly synchronized observations with MASTER II. This instrument consists of a pair of 0.4-m twin telescopes with their polaroids fixed at orthogonal angles: MASTER-IAC II at 0\textdegree/90\textdegree \,  and MASTER-SAAO II at 45\textdegree/135\textdegree. This configuration allows early-time white-band photometry (see Section \ref{sec:MASTER_phot}) and, when there are two sites simultaneously observing the OT, it enables polarization measurements (see Section \ref{sec:MASTER_pol}).

For both MASTER VWF and MASTER II instruments, we use aperture photometry to derive the source flux ({\sc Astropy Photutils}; \citealt{2016ascl.soft09011B}).

\begin{deluxetable*}{ccccccccc}[t!]
\tablecaption{Results of the models applied to GRB 190114C light curves for LT RINGO3 BV/R/I, MASTER VWF/MASTER II r-equivalent and LT IO:O r optical bands. \label{tab:all_alphas}}
\tablecolumns{9}
\tablewidth{0pt}
\tablehead{
\colhead{Band} & \colhead{Instrument} & \colhead{$\alpha _1$} & \colhead{$\alpha _2$}  & \colhead{t} & \colhead{Model} & $\chi^2$/d.o.f   & p-value & \colhead{Figure}\\
\colhead{} & \colhead{} & \colhead{} & \colhead{} & \colhead{(s)} & \colhead{}  & \colhead{} & \colhead{} & \colhead{} }
\startdata
BV & RINGO3 & $1.082 \pm 0.007$ & - & - & PL$^{a}$ & 627/332 & $<0.0001$ & - \\
BV & RINGO3 & $1.49 \pm 0.02$ & $1.005 \pm 0.006$ & $401 \pm 10$ & BPL$^{b}$ & 290/331 & 0.95 & \ref{fig:LC_GRB190114C_RINGO3} \\
r & MASTER + IO:O & $1.33 \pm 0.02$ & - & - & PL & 745/43 &  $<0.0001$ & \ref{fig:LC_GRB190114C_RINGO3} \\
r & MASTER + IO:O & $1.669 \pm 0.013$ & $1.054 \pm 0.011$ & $407_{-19} ^{+20}$ & BPL & 36/42 & 0.72 & \ref{fig:LC_GRB190114C_RINGO3} \\
R & RINGO3 & $1.147 \pm 0.006$ &  - & - & PL & 1432/389 &$<0.0001$ & - \\
R & RINGO3 & $1.575 \pm 0.013$ & $1.040 \pm 0.004$ & $443_{-7} ^{+11}$ & BPL & 345/388 & 0.94 & \ref{fig:LC_GRB190114C_RINGO3} \\
I & RINGO3 & $1.110 \pm 0.008$ & - & - & PL & 2179/365 & $<0.0001$ & - \\
I & RINGO3 & $1.546 \pm 0.011$ & $0.962 \pm 0.005$ & $525_{-12} ^{+11}$ & BPL & 369/364 & 0.41 & \ref{fig:LC_GRB190114C_RINGO3} \\
\hline
BV,r,R,I & MASTER + RINGO3 + IO:O & $2.35 \pm 0.05$ & $0.905 \pm 0.009$ & - & 2 PLs & 1406/1127 & $<0.0001$ & \ref{fig:LC_GRB190114C_RINGO3_RSFS_2} \\
BV,r,R,I & MASTER + RINGO3 + IO:O & $1.711 \pm 0.012$ & $0.707 \pm 0.010$ & $805 \pm 19$, $831 \pm 47$, & PL + BPL & 1174/1123 & 0.14 & \ref{fig:LC_GRB190114C_RINGO3_RSFS_2} \\
& & &  &  $931 \pm 18$, $1083 \pm 20$ & & & &
\enddata
\tablecomments{The first part of Table \ref{tab:all_alphas} includes all the phenomenological models and the second part, the two physical models that relate to a \textquote{reverse plus forward shock} scenario.}
\tablenotetext{a}{PL: power-law}
\tablenotetext{b}{BPL: broken power-law}
\end{deluxetable*}

\subsubsection{MASTER VWF and MASTER II Light Curves} \label{sec:MASTER_phot}

MASTER-SAAO VWF and MASTER-IAC VWF cameras started observations at T$_0+ 30.7\,$s and T$_0+ 38.6\,$s, respectively; by $\sim$T$_0+ 50\,$s, the OT signal-to-noise ratio falls under 5 and the photometry is discarded. We standardize the VWF white band with the r band using 5 stars of $8-10\,$mag from Pan-STARRS DR1 catalogue \citep{2016arXiv161205560C}.

MASTER-SAAO II and MASTER-IAC II observed the OT since $45.9\,$s and $48\,$s post-burst, respectively. Given this instrumental set-up, we align and average the field frames from the two orthogonal polaroid positions and we derive a single photometric measurement per site. Additionally, we apply RINGO3 photometric criterion and we only accept OT detections with signal-to-noise ratios over 20 (see Section \ref{sec:phot_reduc}). We standardize MASTER II white band to the r band using 5 stars of $13-15 \,$mag from Pan-STARRS DR1 catalogue. During MASTER II observations, these stars present a $\sim 0.04\,$mag deviation from the mean. Both MASTER VWF and MASTER II photometry is corrected for mean Galactic extinction A$_{\rm r} = 0.034 \pm 0.001\,$mag \citep{{1998ApJ...500..525S}} and presented in Table \ref{tab:phot} and Figure \ref{fig:LC_GRB190114C_all}.

\subsubsection{MASTER II Polarization Calibration} \label{sec:MASTER_pol}

There have been several lower bound polarization measurements with only one MASTER II site \citep{2016MNRAS.455.3312G, 2017Natur.547..425T}. For GRB 190114C, MASTER-SAAO II and MASTER-IAC II responded to BAT trigger almost simultaneously --- since $\sim 47\,$s post-burst and with an initial temporal lag of $\sim 2.2\,$s --- allowing to completely sample the Stokes plane and measure polarization degree and angle.

To derive the polarization, we first subtract the relative photometric zero-point between MASTER-SAAO and MASTER-IAC observations using field stars. Due to the temporal lag between the two telescopes sites and the fading nature of the source, we also correct for the relative intensity by interpolating over the two time windows. Following RINGO3 calibration, we use \cite{2002A&A...383..360C} method to derive the Stokes q-u parameters, the polarization degree/angle and the confidence levels (see Section \ref{sec:pola_reduc}). We use RINGO3 polarization measurements of CD-27 1309 star (P=$\,0.1-0.3\%$) to subtract MASTER II instrumental polarization (P$\, \sim 7\%$); by doing this, the polarization contribution from the interstellar medium is also removed. During MASTER II observations, CD-27 1309 star shows deviations of $\sim 0.3 \%$ from the mean.

Although the burst is bright at that time, the signal-to-noise and the polarization degree rapidly drop within the first $\sim 100\,$s; we discard observations after $\sim {\rm T}_0 + 200\,$s. Additionally, we derive a lower bound of the polarization degree at $\sim {\rm T}_0 + 73\,$s --- because the 0\textdegree/90\textdegree\, MASTER-IAC II frames were not taken --- using $P_{\rm low} = (I_2-I_1)/(I_1+I_2)$, where $I_1$ and $I_2$ are the source intensity at each ortogonal polaroid position (see \citealt{2016MNRAS.455.3312G,2017Natur.547..425T} for the procedure). In Table \ref{tab:pol}, we present the Stokes parameters and the polarization degree and angle for MASTER II observations. We note that the angle is not calibrated with polarimetric standard stars, which implies that we cannot determine its evolution from MASTER II to RINGO3 observations.

\section{Results} \label{sec:Results}

Here we present the temporal properties of the optical emission (Section \ref{sec:LC_results}), the optical polarization (Section \ref{sec:Pol_results}) and the spectral analysis of the optical and the X-rays emission (Section \ref{sec:SEDs}).

\subsection{The Emission Decay of the Early Optical Afterglow} \label{sec:LC_results}

\begin{figure}[ht!]
\plotone{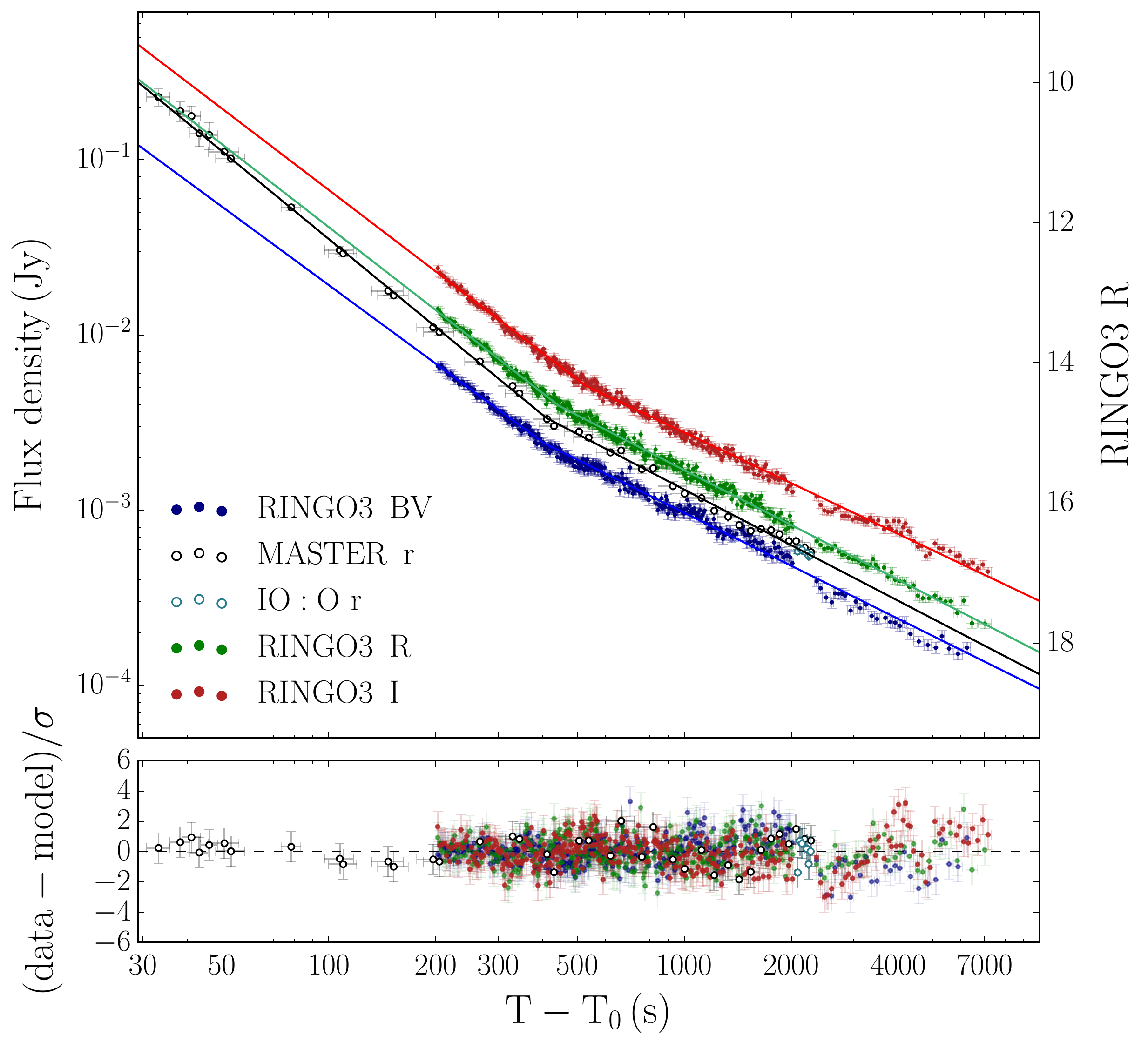}
\caption{GRB 190114C LT and MASTER light curves modeled in terms of broken power-laws:  RINGO3 BV/R/I bands and the joint r-equivalent MASTER VWF/MASTER II/IO:O band. The results of the fits are listed in Table \ref{tab:all_alphas}. The bottom panel corresponds to the residuals of the fit. In the x-axis, T$_0$ corresponds to BAT trigger time; in the y-axis, the flux density is converted to RINGO3 R magnitude.}
\label{fig:LC_GRB190114C_RINGO3}
\end{figure}

A simple power-law model yields a poor fit to the RINGO3 light curves (see Table \ref{tab:all_alphas}). Consequently, we attempt a broken power-law fit to each band, which significantly improves the $\chi^2$ statistics (see Table \ref{tab:all_alphas} and Figure \ref{fig:LC_GRB190114C_RINGO3}). This result indicates a light curve flattening from $\alpha_{\rm opt} \sim 1.5$ to $\alpha_{\rm opt} \sim 1$ at $t_{\rm break, \lbrace BV, R, I \rbrace} = 401 \pm 10 \,$s, $443_{-7} ^{+11}\,$s, $525_{-12} ^{+11}\,$s post-burst. There is a discrepancy between the break times of the three bands that cannot be reconciled within 3$\sigma$, indicating that the break is chromatic and moving redwards through the bands.

A broken power-law model also gives a good fit to the r-equivalent MASTER VWF, MASTER II and IO:O joint light curve (see Table \ref{tab:all_alphas} and Figure \ref{fig:LC_GRB190114C_RINGO3}). Early-time observations from MASTER VWF prove that the optical emission was already decaying as a simple power-law since $30.7\,$s post-burst with $\alpha_{\rm opt} = 1.669 \pm 0.013$. At T$_0 + 407_{-19} ^{+20}\,$s, consistent with RINGO3 BV break time, the light curve flattens to $\alpha_{\rm opt} = 1.054 \pm 0.011$.

\subsection{Time-resolved Polarimetry in White and Three Optical Bands} \label{sec:Pol_results}

\begin{figure}[ht!]
\plotone{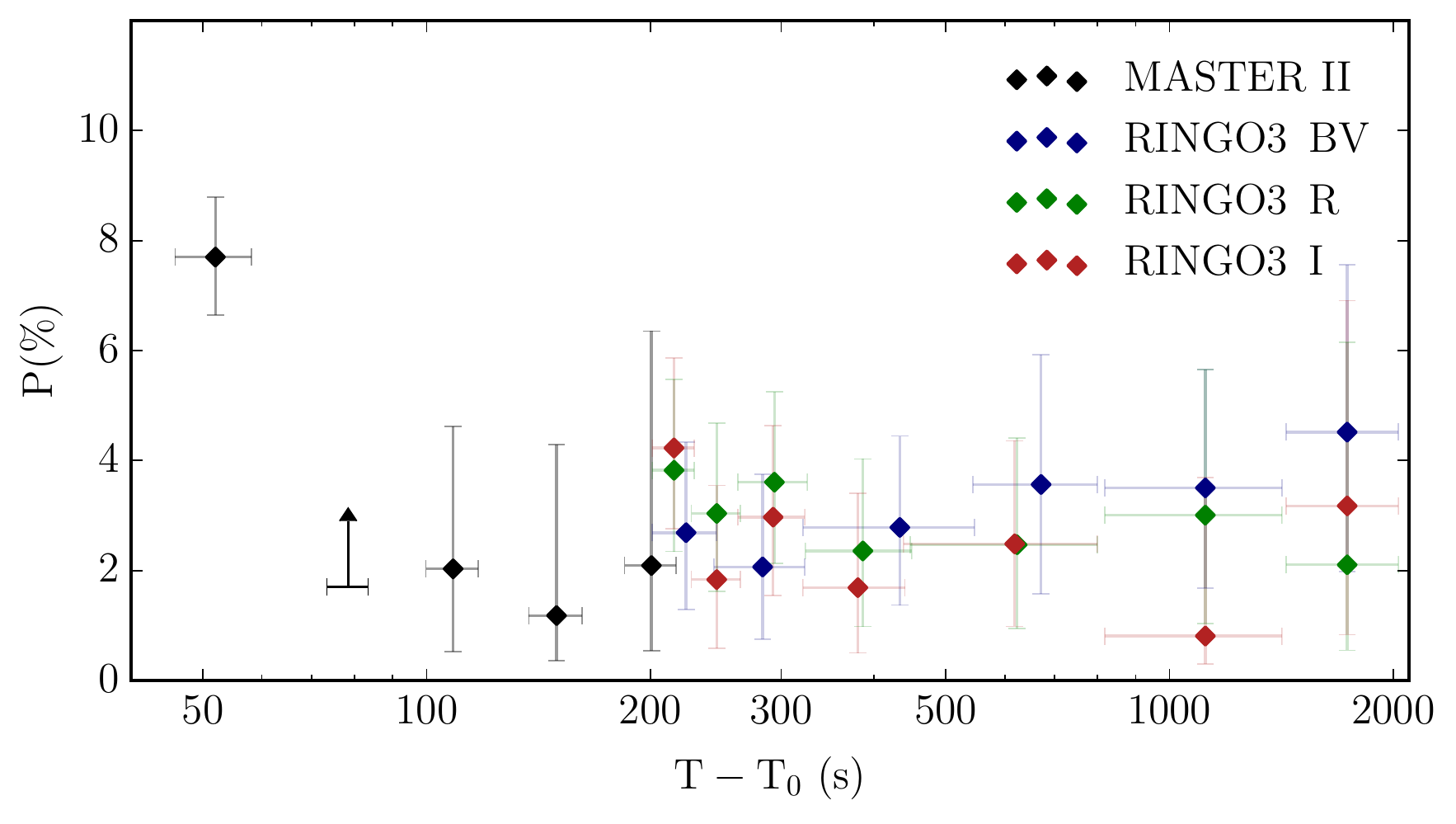}
\caption{MASTER II and RINGO3 temporal evolution of the polarization degree. Uncertainties are quoted at $2\sigma$ confidence level. The black arrow corresponds to a $2\sigma$ lower bound of the polarization degree.}
\label{fig:P_evolution_all}
\end{figure}

During the first $\sim 50\, $s of MASTER II observations, the polarization degree displays an early-time drop from $7.7 \pm 1.1 \%$ to $2.0 ^{+2.6} _{-1.5}$ consistent with the constant low polarization degree measured by RINGO3 from $\sim 200\,$s onwards (see Figure \ref{fig:P_evolution_all}). From $52\,$s to $109\,$s post-burst, the polarization angle remains constant within uncertainties (see Table \ref{tab:pol}).

\begin{figure*}[t!]
\plottwo{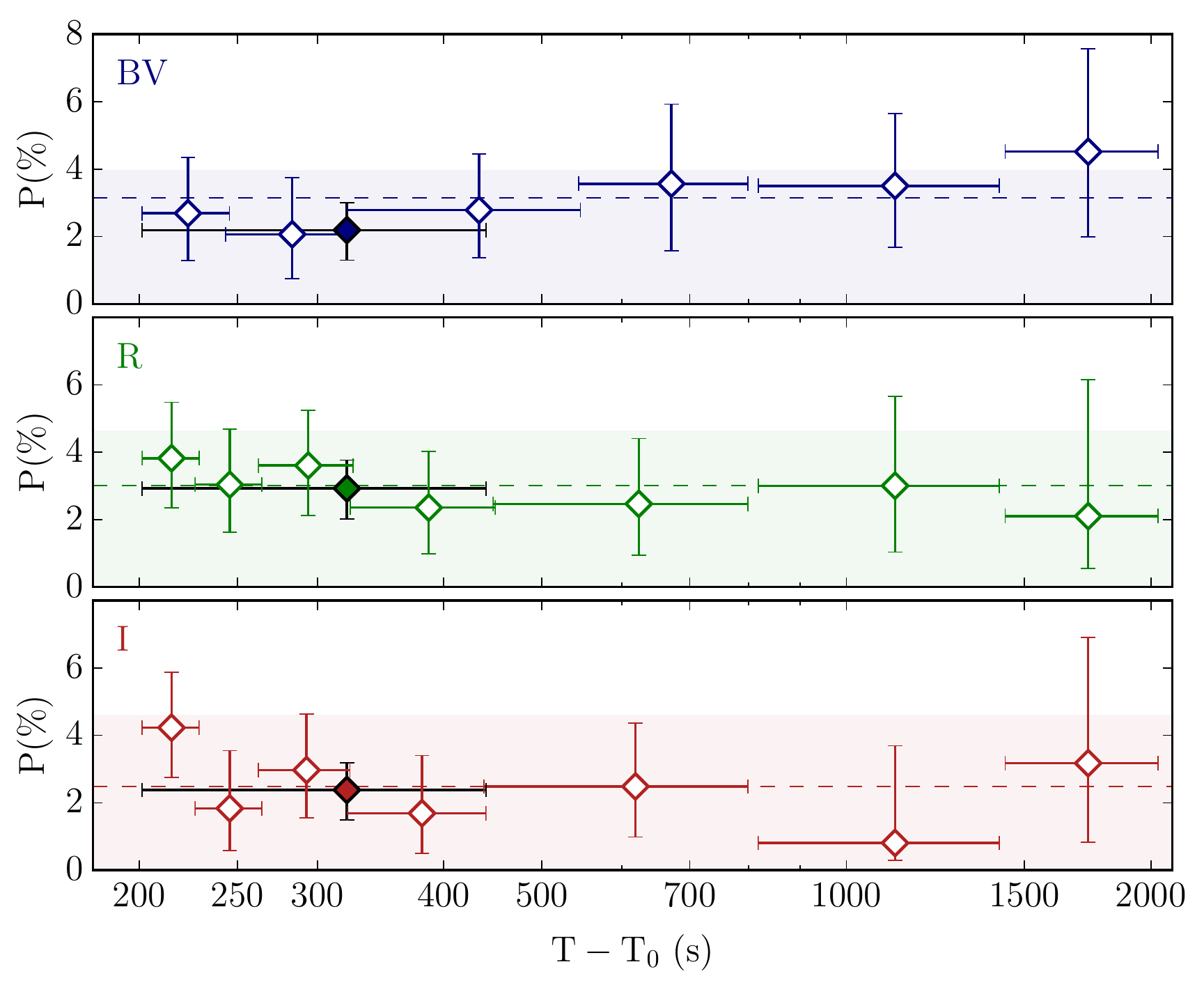}{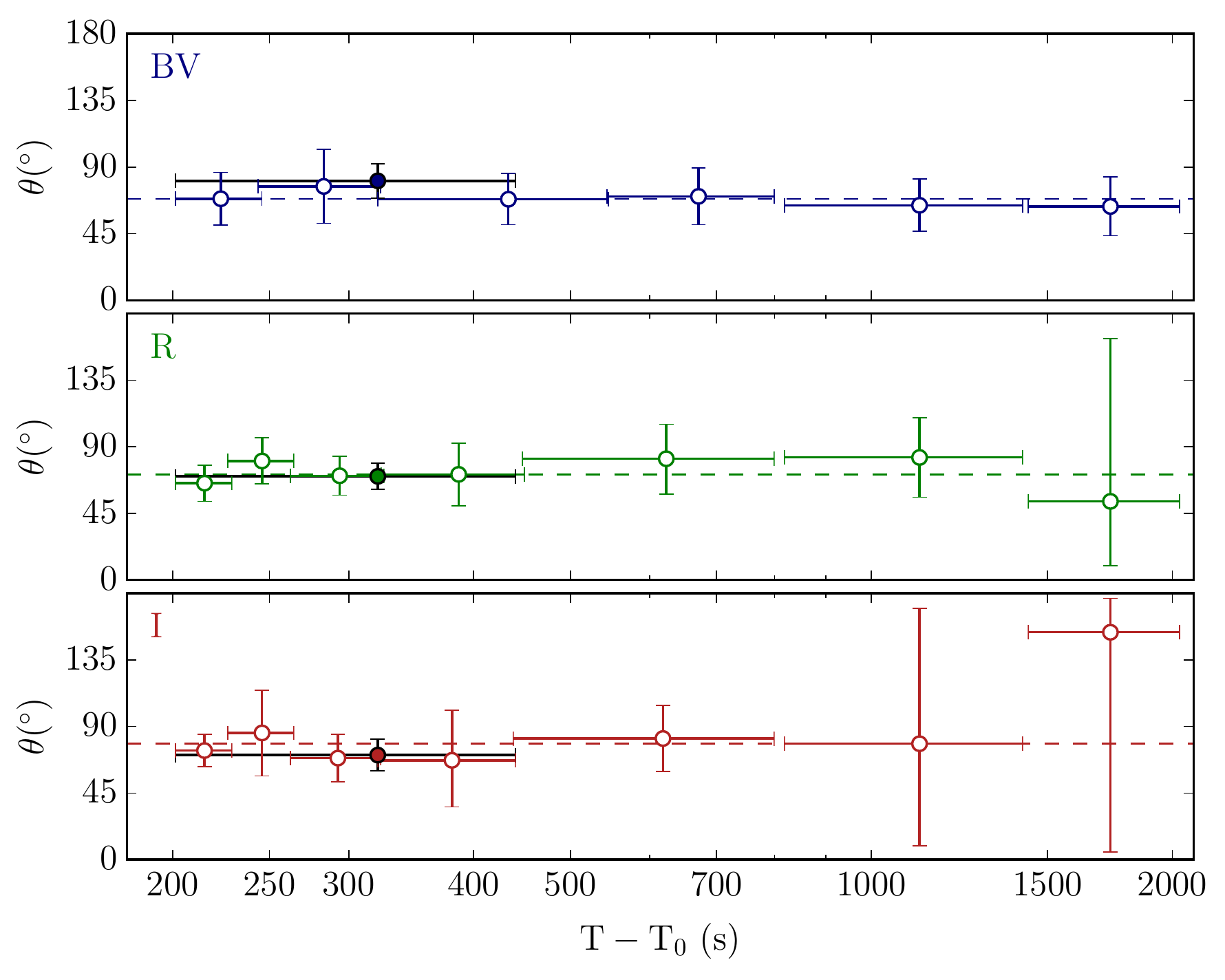}
\caption{GRB 190114C temporal evolution of the polarization degree (left) and angle (right) for the three RINGO3 bands. In black, we show the maximum signal-to-noise integration. Uncertainties are quoted at $2\sigma$ confidence level. Dotted lines correspond to the median value and the shaded region to the maximum polarization degree induced by dust in the line-of-sight: including the highly extincted host galaxy and a small contribution from the MW (E$_{\rm B-V, HG}=0.51 \pm 0.04$; E$_{\rm B-V, MW}=0.0124 \pm 0.0005$). T$_0$ corresponds to BAT trigger time.}
\label{fig:P_evolution}
\end{figure*}

RINGO3 time-resolved polarization show constant degree and angle within $2\sigma$ confidence level during $\sim 200-2000\, $s post-burst (see Figure \ref{fig:P_evolution}), ruling out any temporal trend at these timescales or swings in polarization bigger than $\Delta $P$ \sim 3\%$ for $t \sim 200-450\,$s post-burst and at $3\sigma$ confidence level. The temporal behavior of polarization agrees with the value inferred in Section \ref{sec:pola_reduc} from the maximum signal-to-noise integration: P$_{\rm \lbrace BV, R, I \rbrace}= 2.2_{-  0.8} ^{+  0.9} \%$, $2.9_{-  0.8} ^{+  0.9} \%$, $2.4_{-  0.8} ^{+  0.9} \%$, $\theta_{\rm \lbrace BV, R, I \rbrace}= 81 \pm 12$\textdegree, $70 \pm 9$\textdegree, $71 \pm 11 $\textdegree (see Figure \ref{fig:P_evolution} black observations) and the median value: P$_{\rm \lbrace BV, R, I \rbrace}= 3.1 \pm 0.4 \%, 3.0 \pm 0.6 \%, 2.5 \pm 0.7 \%$, $\theta_{\rm \lbrace BV, R, I \rbrace} = 68 \pm 3 $\textdegree, $71 \pm 9$\textdegree, $78 \pm 7$\textdegree (quoting the median absolute deviation; see Figure \ref{fig:P_evolution} doted lines). The behavior is the same in all three bands.

\subsection{The Spectral Evolution of the Afterglow} \label{sec:SEDs}

\begin{figure}[ht!]
\plotone{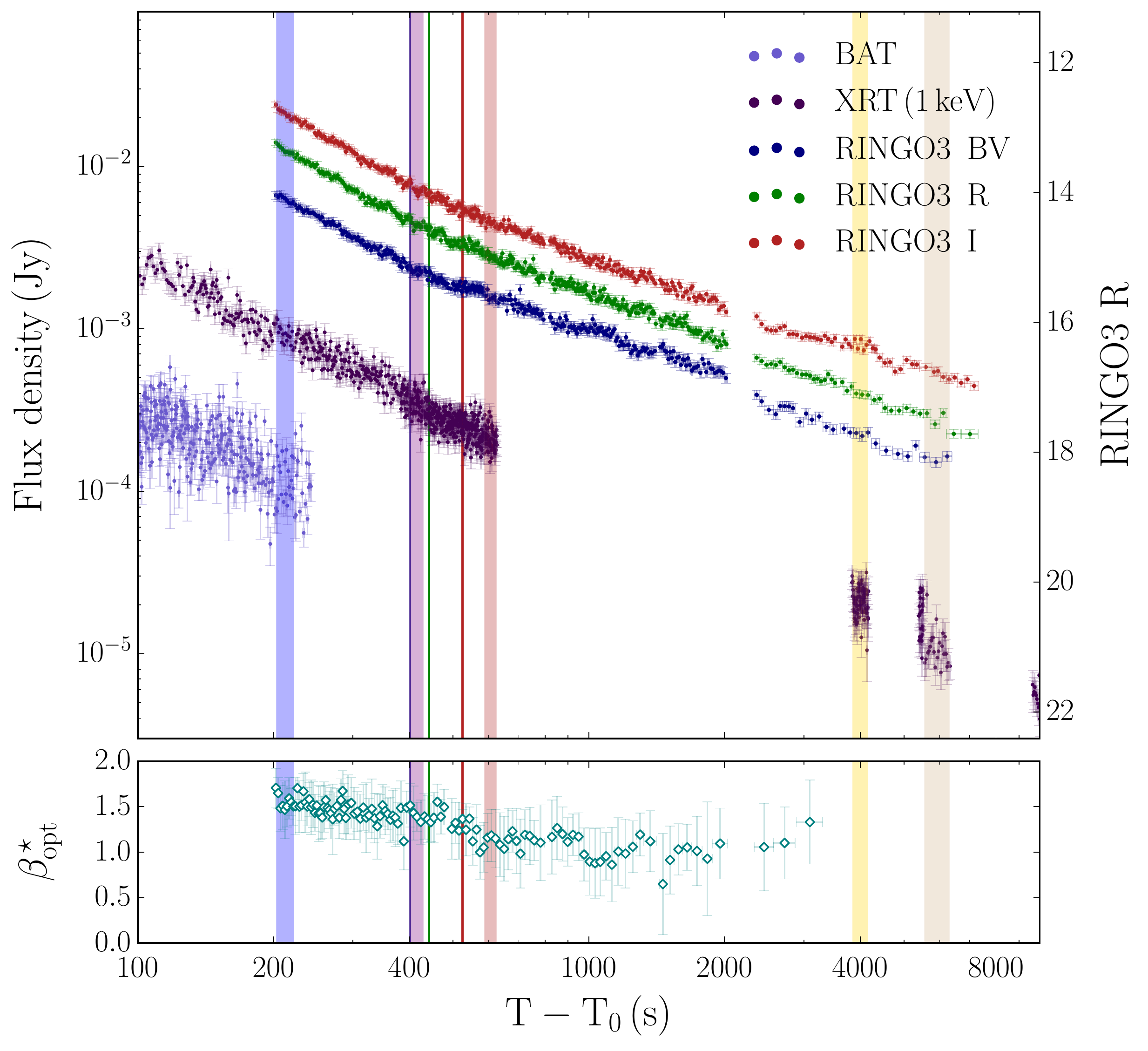}
\caption{GRB 190114C BAT/XRT \citep{2009MNRAS.397.1177E} and RINGO3 BV/R/I light curves with the observations used for the broadband spectral energy distribution modeling highlighted in shaded colors. The vertical solid lines correspond to RINGO3 BV/R/I light curves break times. The bottom panel corresponds to the optical spectral index inferred from RINGO3 BV/R/I bands without considering host galaxy extinction. In the x-axis, T$_0$ corresponds to BAT trigger time; in the y-axis, the flux density is converted to RINGO3 R magnitude.}
\label{fig:LC_GRB190114C_zoomin_all}
\end{figure}

To spectrally characterize GRB 190114C during RINGO3 observations, we test for color evolution in the optical (Section \ref{sec:SEDs_color}), we study the spectral evolution of the 0.3-150 keV X-rays band for the time-intervals of Figure \ref{fig:LC_GRB190114C_zoomin_all} (Section \ref{sec:SEDs_xrays}) and we check how the optical and the X-rays connect (Section \ref{sec:SEDs_broadband}).

\subsubsection{Color Evolution through RINGO3 Bands} \label{sec:SEDs_color}

Taking advantage of the simultaneity of RINGO3 three-band imaging, we attempt to infer the evolution of the optical spectral index. To guarantee a spectral precision of $\sim 0.05-0.06\, $mag per measurement, we take the lowest signal-to-noise light curve (BV band) and we dynamically co-add frames so the OT reaches a signal-to-noise threshold of $\geq 40$. Following, we co-add R/I frames using the BV band binning and for every three-band spectral energy distribution (SED), we fit a power-law.

In Figure \ref{fig:LC_GRB190114C_zoomin_all}, we present the evolution of the optical spectral index $\beta_{\rm opt} ^{\, \star}$; this index is not corrected for host galaxy extinction (see Section \ref{sec:SEDs_broadband}), which makes this measurement an upper limit of the intrinsic $\beta_{\rm opt}$. Spectral indexes exhibit a decreasing behavior from $\beta_{\rm opt} ^{\, \star} \sim 1.5$ to $\beta_{\rm opt} ^{\, \star} \sim 1$ masked by the uncertainties. Due to the number of measurements available, we perform a Wald-Wolfowitz runs test \citep{wald1940} of all the points against the median value to check for a trend. If there is no real decrease of the spectral index, the data should fluctuate randomly around the median. In this case, a run is a consecutive series of $\beta_{\rm opt} ^{\, \star}$ terms over or under the median. The temporal evolution of the spectral indexes displays significantly smaller number of runs than expected with p-value$\,=2 \times 10^{-15}$, which rejects the hypothesis of randomness and indicates that a temporal trend from soft to harder spectral indexes is likely. This result is in agreement with the chromatic nature of the break observed in the RINGO3 light curves.

\subsubsection{The 0.3-150 keV X-rays Spectra}  \label{sec:SEDs_xrays}

For the X-rays spectral analysis, we use the available BAT-XRT observations that correspond to the time-intervals of Figure \ref{fig:LC_GRB190114C_zoomin_all}. With this choice, the first spectrum is before the slope change of the optical light curve at $\sim 400-500\, $s post-burst (see Section \ref{sec:LC_results}). Due to the synchrotron nature of the afterglow, the models used for this analysis comprise either a single power-law or connected power-laws. 

We extract the time-resolved 0.3-10 keV XRT spectra using the web interface provided by Leicester University\footnote{http://www.swift.ac.uk/user\_objects/} based on {\sc heasoft} (v. 6.22.1; \citealt{1995ASPC...77..367B}). Energy channels are grouped with {\sc grppha} tool so we have at least 20 counts per bin to ensure the Gaussian limit and adopt $\chi^2$ statistics. The first four time-intervals were observed in WT mode and the final one in PC mode. For modeling WT observations, we only consider energies $\geq 0.8 \,$keV due to an instrumental effect that was reported in \cite{2019GCN.23736....1B}. Simultaneous time-resolved, 15-150 keV spectra with BAT are extracted for the first three time-intervals using the standard BAT pipeline (e.g., see \citealt{Rizzuto07}) and are finally grouped in energy to ensure a $>2\sigma$ significance. 

\begin{figure}[ht!]
\plotone{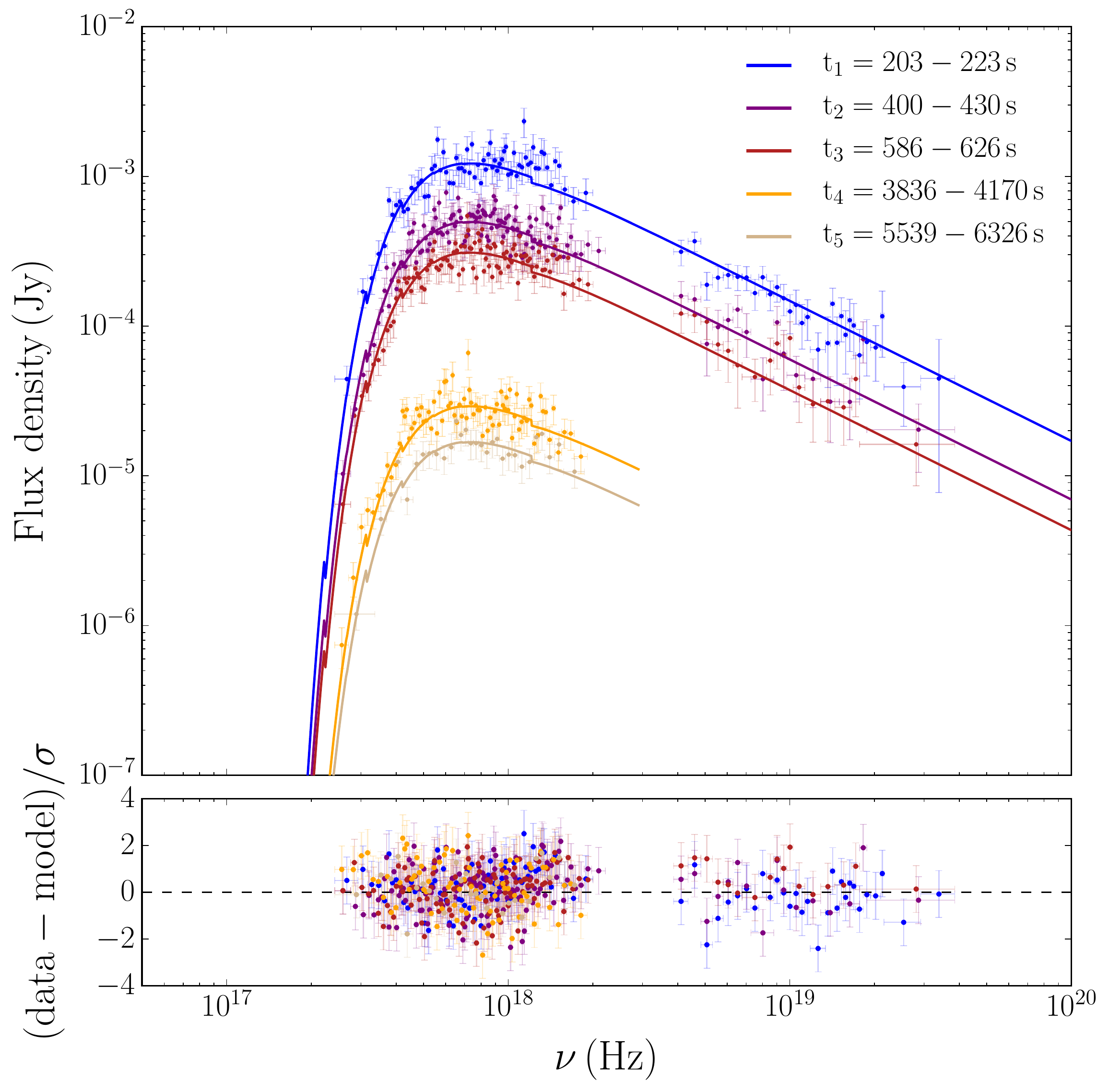}
\caption{GRB 190114C X-rays spectra of the combined 0.3-10 keV XRT and 15-150 keV BAT observations \citep{2009MNRAS.397.1177E}. The spectra are modeled with an absorbed power-law that accounts for the Galactic and host galaxy rest-framed total hydrogen absorption. The results of the fit are: $\beta_{\rm x}=0.94 \pm 0.02$, N$_{\rm H, HG}= (9.3 \pm 0.2) \times 10^{22}\,$cm$^{-2}$ with $\chi^2 /{\rm d.o.f.}= 422/466$ and p-value$\,\,= 0.89$. The bottom panel corresponds to the residuals of the fit.}
\label{fig:SED_xrays}
\end{figure}

The combined BAT-XRT spectra are modeled under {\sc xspec} (v. 12.9.1; \citealt{1999ascl.soft10005A}) using $\chi^2$ statistics with a simple absorbed power-law ({\sc powerlaw*phabs*zphabs}) that accounts for the rest-framed host galaxy total hydrogen absorption, N$_{\rm H, HG}$, and the Galactic\footnote{Derived using https://www.swift.ac.uk/analysis/nhtot/ tool} N$_{\rm H, MW}=  7.54 \times 10^{19} \,{\rm cm}^{-2}$ \citep{2013MNRAS.431..394W}. By satisfactorily fitting each spectra with a power-law, we find that the 0.3-10 keV and 15-150 keV spectra belong to the same spectral regime and that there is no significant spectral evolution during the first $\sim 200-6000\,$s post-burst. In Figure \ref{fig:SED_xrays}, we fit all five spectra with a single spectral index. The fit procedure results in an spectral index $\beta_{\rm x}=0.94 \pm 0.02$, rest-frame hydrogen absorption N$_{\rm H, HG}= (9.3 \pm 0.2)  \times 10^{22}\,$cm$^{-2}$, $\chi^2 /{\rm d.o.f.}= 422/466$ and p-value$\,\,= 0.89$. Due to the high column density absorption among the soft X-rays, the slope is mainly constrained by the hard X-rays.

\subsubsection{Broadband Spectral Energy Distributions} \label{sec:SEDs_broadband}

We obtain the combined BAT-XRT-RINGO3 spectral energy distributions (SEDs) by co-adding those RINGO3 frames that correspond to a given X-rays epoch and then deriving the absolute flux calibrated photometry (see Section \ref{sec:RINGO3bandpasses}).

Broadband SEDs are also modeled under {\sc xspec} using $\chi^2$ statistics with a simple absorbed power-law ({\sc powerlaw*zdust*zdust*phabs*zphabs}) that accounts for total hydrogen absorption (see Section \ref{sec:SEDs_xrays}), Galactic extinction (E$_{\rm B-V, MW} = 0.0124 \pm 0.0005$; \citealt{1998ApJ...500..525S}) and a rest-framed SMC dust extinction profile for the host galaxy \citep{1992ApJ...395..130P}.

\begin{figure}[ht!]
\plotone{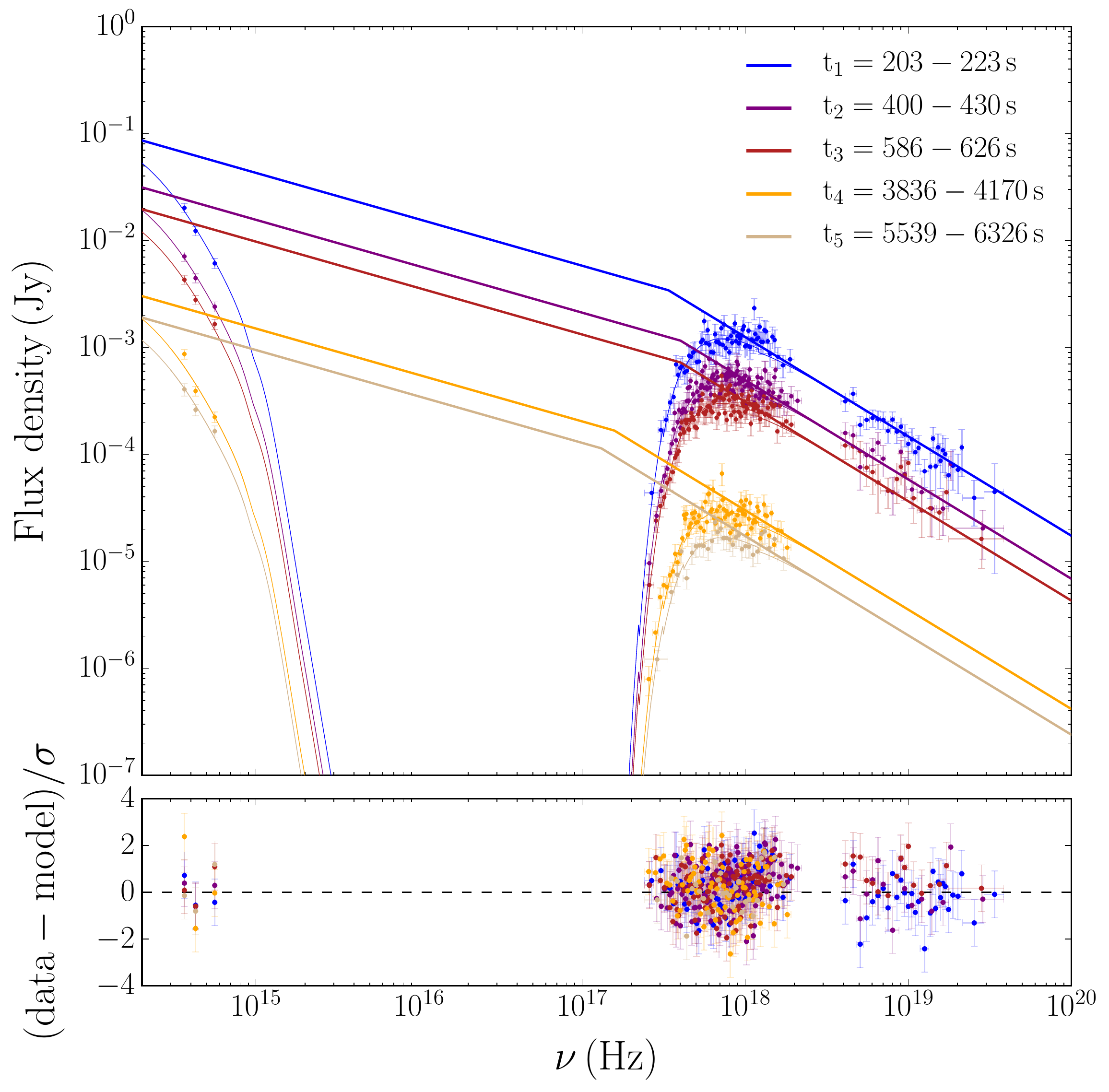}
\caption{GRB 190114C broadband SEDs with RINGO3, XRT and BAT observations \citep{2009MNRAS.397.1177E}. SEDs are best fitted with a broken power-law model that accounts for extinction in the optical and total hydrogen absorption in the X-rays. The results of the fit are: $\beta_{\rm opt}=0.43 \pm 0.02$, $\beta_{\rm x}=0.93 \pm 0.02$, E$_{\rm break, \, \lbrace 1,2,3,4,5 \rbrace}  = 1.4 \pm 0.3\,$keV, $1.6 \pm 0.2 \,$keV, $1.6 \pm 0.3\,$keV, $0.65 \pm 0.14\,$keV, $0.54 \pm 0.12\,$keV with $\chi^2/{\rm d.o.f.}=439/481$ and p-value$\,\,= 0.82$; in the host galaxy rest-frame: A$_{\rm v, HG} = 1.49 \pm 0.12 \,$mag and N$_{\rm H, HG}= (9.0 \pm 0.3) \times 10^{22}\,$cm$^{-2}$. The bottom panel corresponds to the residuals of the fit.}
\label{fig:SED_optical}
\end{figure}

\begin{deluxetable*}{ccccccccc}[t!]
\tablecaption{Optical and X-rays temporal $\alpha$ and spectral $\beta$ indexes of GRBs
with optical light curves that show a steep-to-flat behavior and decay rates comparable to GRB 190114C.
 \label{tab:grbs_steep_to_flat_LC}}
\tablecolumns{8}
\tablewidth{0pt}
\tablehead{
\colhead{GRB} & \colhead{$\alpha _{\rm opt, 1}$} & \colhead{$\alpha _{\rm opt, 2}$}  & \colhead{$\alpha _{\rm x}$} & \colhead{$\beta _{\rm opt}$}  & \colhead{$\beta _{\rm x}$} & \colhead{Reference} \\
\colhead{} & \colhead{} & \colhead{} & \colhead{} & \colhead{} & \colhead{} & \colhead{} }
\startdata
021211 & $\sim 1.6$ & $\sim 1.1$ & - & $\leq 0.98$ & -  & \cite{2003ApJ...586L...5F} \\
050525A & $\sim 1.3$ & $\sim 1$ & $0.68 ^{+0.06} _{-2.18}-1.54 \pm 0.06$ & - & $0.97^{+0.16} _{-0.15}$ & \cite{2005ApJ...633.1027S,2009MNRAS.397.1177E} \\
050904 & $1.36 ^{+0.07} _{-0.06}$ & $0.82^{+0.21} _{-0.08}$ & $ 2.02^{+0.06} _{-0.05} - 1.39^{+0.06} _{-0.05}$ & $\leq −1.25 ^{+0.15} _{-0.14}$ & $0.84 ^{+0.06} _{-0.05}$ & \cite{2006Natur.440..181H,2009MNRAS.397.1177E} \\
060908 & $1.5 \pm 0.3$ & $1.05 \pm 0.03$ & $1.14 ^{+0.03} _{-0.02}$ & $\sim 0.3$ & $1.1 \pm 0.2$ & \cite{2010AandA...521A..53C, 2009MNRAS.397.1177E} \\
061126 & $1.48 \pm 0.06$ & $0.88 \pm 0.03$ & $1.290 \pm 0.008$ & $0.38 \pm 0.03^{\rm a}$ & $0.88 \pm 0.03$ & \cite{2008ApJ...687..443G} \\
090102 & $1.50 \pm 0.06$ & $0.97 \pm 0.03$ & $1.34 \pm 0.02$ & $\leq 1.32$ & $0.83 \pm 0.09$ & \cite{2010MNRAS.405.2372G} \\
090424 & $\sim 1.5$ & $\sim 0.85$ & $0.87 \pm 0.02-1.17 \pm 0.01$ & - & $0.87 \pm 0.09$ & \cite{2013ApJ...774..114J, 2009MNRAS.397.1177E} \\
090902B & $\sim 1.6$ & $0.90 \pm 0.08$ & $1.30 \pm 0.04$ & $0.68 \pm 0.11$ & $0.9 \pm 0.1$ & \cite{2010ApJ...714..799P} \\
190114C & $1.669 \pm 0.013$ & $\sim 1$ & $1.345 \pm 0.004$ & $0.43 \pm 0.02$ & $0.93 \pm 0.02$ & This work
\enddata
\tablenotetext{a}{$\beta _{\rm x}$ is linked to $\beta _{\rm opt}$ as $\beta _{\rm x} = \beta _{\rm opt} + 0.5$.}
\end{deluxetable*}

The optical and X-ray fluxes do not connect with a simple absorbed power law. Consequently, we test for a break between the two spectral regimes (using {\sc bknpower} model). For all five SEDs, we link all parameters relating to absorption, extinction and spectral indexes and we leave the break frequency as a free parameter for each SED. From the broken power-law fit (see Figure \ref{fig:SED_optical}), we obtain a spectral index $\beta_{\rm opt}=0.43 \pm 0.02$ for the optical and $\beta_{\rm x}=0.93 \pm 0.02$ for the X-rays with $\chi^2 /{\rm d.o.f.} = 439/481$ and p-value$\,\,= 0.82$. The break evolves as E$_{\rm break, \, \lbrace 1,2,3,4,5 \rbrace}  = 1.4 \pm 0.3\,$keV, $1.6 \pm 0.2\,$keV, $1.6 \pm 0.3\,$keV, $0.65 \pm 0.14\,$keV, $0.54 \pm 0.12\,$keV. We derive high extinction A$_{\rm v, HG} = 1.49 \pm 0.12 \,$mag, or equivalently, E$_{\rm B-V, HG}=0.51 \pm 0.04$, and absorption N$_{\rm H, HG}= (9.0 \pm 0.3) \times 10^{22}\,$cm$^{-2}$ at the host galaxy rest-frame. We achieve compatible results within 1$\sigma$ for spectral indexes, energy breaks and total hydrogen absorption using LMC/MW dust extinction profiles, which gives A$_{\rm v, HG} =1.64 \pm 0.13 \,$mag, $1.72 \pm 0.12 \,$mag, respectively.

\section{Theoretical Modeling} \label{sec:th_model}

\subsection{Modeling the Optical Afterglow} \label{sec:after_model}

In the standard fireball model, possible mechanisms that produce chromatic breaks include the passage of a break frequency through the band, a change in the ambient density profile or an additional emission component \citep{2008ApJ...686.1209M}. We rule out that the light curve flattening at $\sim 400-500\,$s post-burst and at magnitude $\sim 14$ is due to an emerging supernova --- \cite{2019GCN.23983....1M} reported a supernova component $15\,$days post-burst --- or host galaxy contamination. Additionally, optical emission from ongoing central engine activity is unlikely: BAT/XRT emission is already decaying since $\sim 30\,$s and $\sim 70\,$s post-burst, respectively (see Figure \ref{fig:LC_GRB190114C_all}).

Several GRBs exhibit a similar light curve flattening from $\alpha_{\rm opt, 1} \sim 1.3-1.7$ to $\alpha_{\rm opt, 2} \sim 0.8-1.1$ in the optical at early times; see Table \ref{tab:grbs_steep_to_flat_LC}: GRB 021211 \citep{2003ApJ...586L...5F}, GRB 050525A \citep{2005ApJ...633.1027S}, GRB 050904 \citep{2006Natur.440..181H,2006ApJ...636L..69W}, GRB 060908 \citep{2010AandA...521A..53C}, GRB 061126 \citep{2008ApJ...687..443G, 2008ApJ...672..449P}, GRB 090102 \citep{2009Natur.462..767S, 2010MNRAS.405.2372G}, GRB 090424 \citep{2013ApJ...774..114J} and GRB 090902B \citep{2010ApJ...714..799P}. Additionally, most of them bear similar spectral and temporal properties to GRB 190114C in both optical and X-rays regimes.

For GRB 021211, GRB 050525A, GRB 061126, GRB 090424 and GRB 090902B, the optical excess at the beginning of the light curve favored the presence of reverse shock emission \citep{2003ApJ...586L...5F, 2005ApJ...633.1027S, 2008ApJ...687..443G,2008ApJ...672..449P,2010ApJ...714..799P, 2013ApJ...774..114J}. Due to a quasi-simultaneous X-rays and optical flare, GRB 050904 light curve was better understood in terms of late-time internal shocks \citep{2006ApJ...636L..69W}. For GRB 090102, \cite{2010MNRAS.405.2372G} also considered the possibility of a termination shock caused by a change in the surrounding medium density profile. However, \cite{2009Natur.462..767S} $10 \pm 1 \%$ polarization measurement during the steep decay of the afterglow favored the presence of large-scale magnetic fields and therefore, of a reverse shock component. Additionally, \citealt{2013Natur.504..119M} reported $28 \pm 4 \%$ polarization degree at the peak of GRB 120308A optical emission, a decline to $16 ^{+5} _{-4} \%$ and a light curve flattening which was interpreted as a reverse-forward shock interplay. Therefore, we attempt to model GRB 190114C optical emission with a reverse plus forward shock model.

\subsubsection{Reverse-Forward Shock Model}\label{sec:reverse}

Under the fireball model framework, the evolution of the spectral and temporal properties of the afterglow satisfy closure relations \citep{1998ApJ...497L..17S, 2004IJMPA..19.2385Z,2006ApJ...642..354Z, 2009ApJ...698...43R, 2013NewAR..57..141G}. These depend on the electron spectral index $p$, the density profile of the surrounding medium (ISM or wind), the cooling regime (slow or fast) and the jet geometry. In the reverse shock scenario, the total light curve flux can be explained by a two-component model that combines the contribution of the reverse and forward shock emission \citep{2000ApJ...545..807K, 2003ApJ...582L..75K, 2003ApJ...595..950Z}. 

\begin{figure*}[ht!]
\plottwo{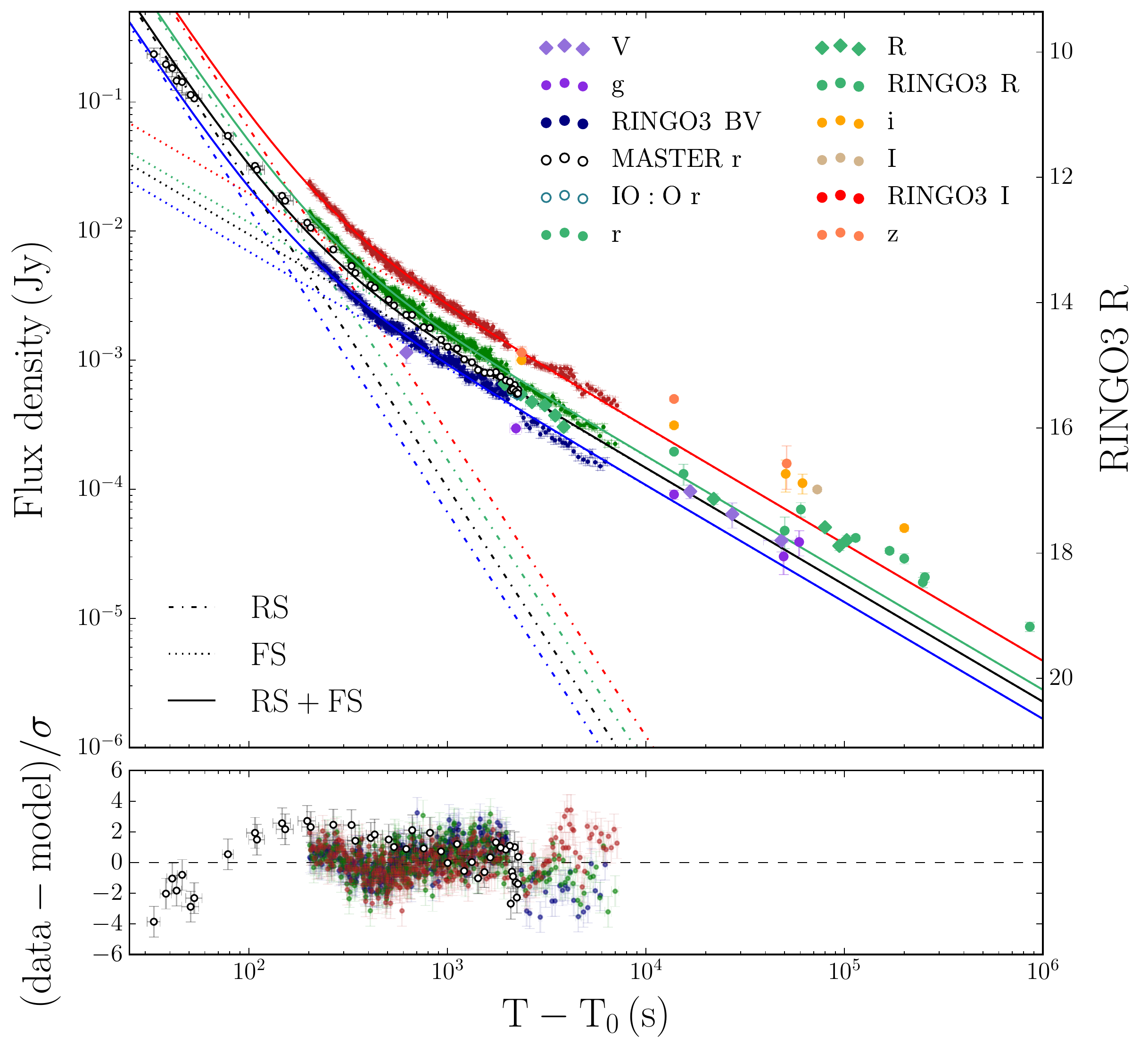}{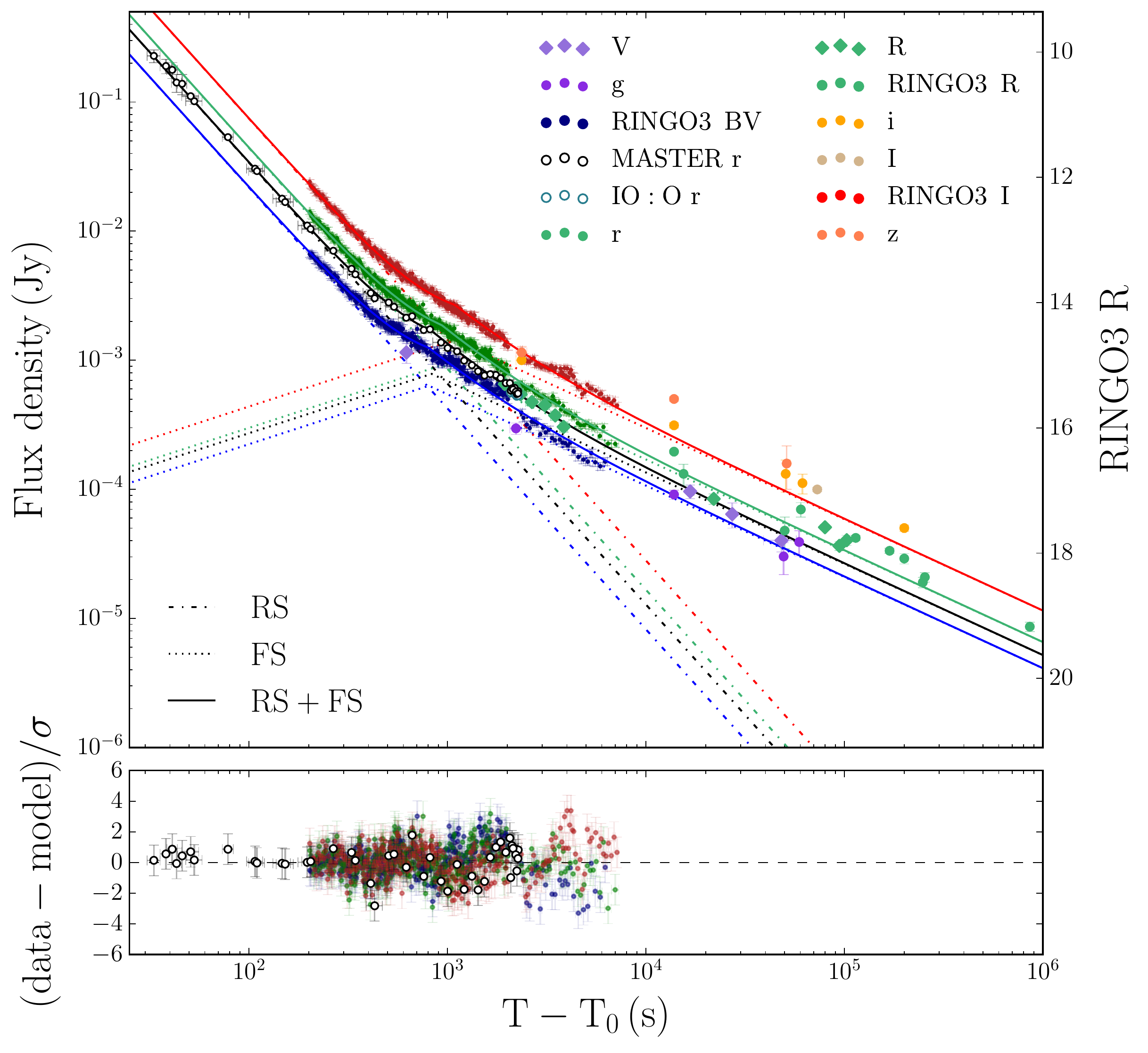}
\caption{GRB 190114C MASTER/IO:O r-equivalent and RINGO3 BV/R/I  multi-wavelength light curves modeled in terms of reverse (RS) plus forward shock (FS) emission. On the left, we model the two components in terms of power-laws. On the right, the forward shock peaks during observations with a fixed $\alpha$ rise of $0.5$ (expected for ISM, slow cooling regime and for the spectral configuration $\nu_{\rm m,f} < \nu_{\rm opt}<\nu_{\rm c,f}$). The results of both fits are listed in Table \ref{tab:all_alphas}; the bottom panels correspond to the residuals of the fits. We also display the data reported in GCNs that cover energy ranges similar to RINGO3 bandpasses: UVOT \citep{2019GCN.23725....1S}, NOT \citep{2019GCN.23695....1S}, OASDG \citep{2019GCN.23699....1I}, GROND \citep{2019GCN.23702....1B}, REM \citep{2019GCN.23754....1D}, McDonald observatory \citep{2019GCN.23717....1I}, LSGT \citep{2019GCN.23732....1K}, GRowth-India \citep{2019GCN.23733....1K}, KMTNet \citep{2019GCN.23734....1K}, UKIRT \citep{2019GCN.23757....1I}, CHILESCOPE \citep{2019GCN.23741....1M, 2019GCN.23746....1M, 2019GCN.23787....1M}, RTT150 \citep{2019GCN.23766....1B}, ePESSTO NTT \citep{2019GCN.23748....1R}, RATIR \citep{2019GCN.23751....1W} and HCT \citep{2019GCN.23742....1K,2019GCN.23798....1S}. GCNs observations do not include filter corrections. In the x-axis, T$_0$ corresponds to BAT trigger time; in the y-axis, the flux density is converted to RINGO3 R magnitude.}
\label{fig:LC_GRB190114C_RINGO3_RSFS_2}
\end{figure*}

The reverse shock emission produces a bright optical peak when the fireball starts to decelerate at $t_{\rm peak, r}$, which happened prior to the MASTER/RINGO3 observations. For ISM, slow cooling regime and with the optical band in between the typical synchrotron and cooling frequency, $\nu_{\rm m,r} < \nu_{\rm opt}<\nu_{\rm c,r}$, the emission should decay\footnote{The decay rate is much slower or faster if the observations are in another spectral regime or/and the emission is due to high latitude emission \citep{2000ApJ...545..807K, 2003ApJ...597..455K}} with $\alpha_{\rm r}=(3p+1)/4 \sim 2$ for a typical $p \sim 2.3$. Later on, the forward shock peaks when the typical synchrotron frequency $\nu_{\rm m, f}$ crosses the optical band. In the $\nu_{\rm m,f} < \nu_{\rm opt}<\nu_{\rm c,f}$ spectral regime, the forward shock emission will follow an expected decay with $\alpha_{\rm f}= 3(p-1)/4 \sim 1$, which flattens the light curve. Consequently, the reverse-forward shock model consists of a power-law with a temporal decay $\alpha_{\rm r}$ for the reverse shock component plus a forward shock contribution that has an expected rise 0.5 and decay $\alpha_{\rm f}$. GRB 190114C light curves suggest that the forward shock peak time $t_{\rm peak, f}$ happens before or during MASTER/RINGO3 observations --- masked by the bright reverse shock emission.

In the left panel of Figure \ref{fig:LC_GRB190114C_RINGO3_RSFS_2}, we attempt the simplest model by considering that the forward shock peaks before MASTER observations ($t_{\rm peak, r},  t_{\rm peak, f} \ll 30\,$s). We leave the reverse and forward shock electron indexes as free parameters. The light curve is best modeled with two power-law components that decay as $\alpha_{\rm opt, r} =2.35 \pm 0.05$ and $\alpha_{\rm opt, f} = 0.905 \pm 0.009$ (see Table \ref{tab:all_alphas}). However, MASTER residuals present a trend and the model underestimates by $\sim 0.8\,$mag late-time observations in the r band reported in GCNs; a decay of $\sim 0.7-0.8$ was reported by \cite{2019GCN.23733....1K} and \cite{2019GCN.23798....1S} hours to days post-burst, which is inconsistent with the $\alpha_{\rm opt, f}$ derived. In addition, UVOT white band emission is decaying as $\alpha=1.62 \pm 0.04$ since $\sim 70 \,$s post-burst with a change to $\alpha=0.84 \pm 0.02$ at $\sim 400\,$s \citep{2019arXiv190910605A}.

In the right panel of Figure \ref{fig:LC_GRB190114C_RINGO3_RSFS_2}, we consider a model in which the forward shock peaks during MASTER/RINGO3 observations. In this model, the two emission components decay as $\alpha_{\rm opt, r} = 1.711 \pm 0.012$, $\alpha_{\rm opt, f} = 0.707 \pm 0.010$ and the forward shock peaks at $t_{\rm peak, f, \lbrace BV, r, R, I \rbrace } = 805 \pm 19 \,$s, $831 \pm 47\,$s, $931 \pm 18\,$s, $1083 \pm 20 \,$s (see Table \ref{tab:all_alphas}). Both reverse and forward shock decay indexes are compatible with an electron index $p \sim 1.95$. Allowing different peak times for each band is preferred over a fixed peak time model; consistent with a chromatic emergence of the forward shock that moves redwards through the bands. The typical synchrotron break frequency is expected to evolve through RINGO3 bands like $\nu_{\rm m, f} \propto t^{-\alpha _{\rm m} }$ with $\alpha _{\rm m} = 1.5$; we find $\alpha_{\rm m} \sim 1.4$.

\begin{figure}[ht!]
\plotone{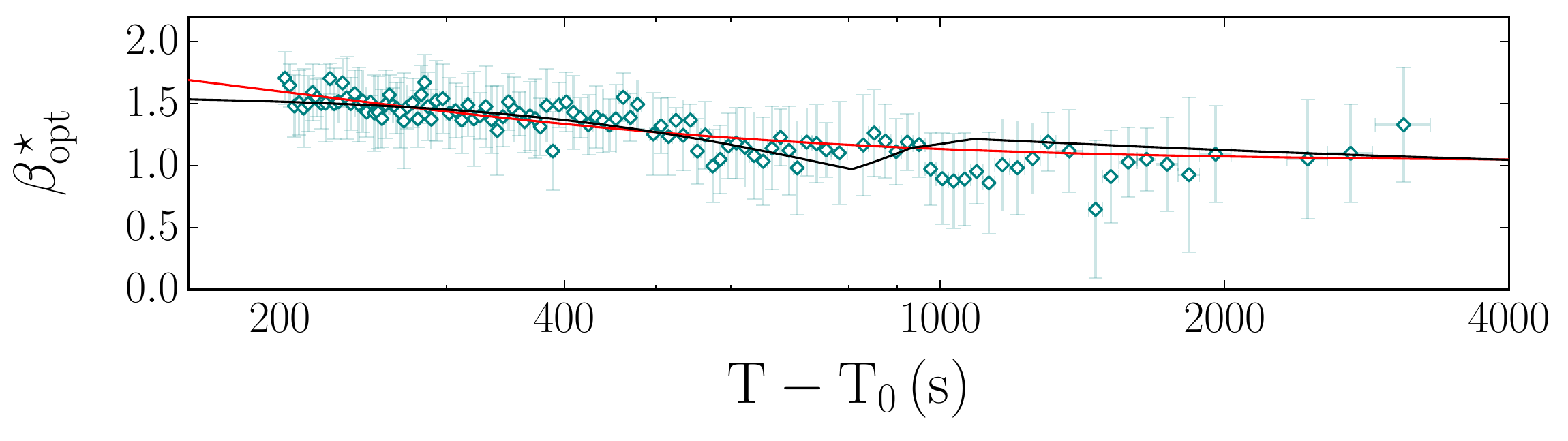}  
\caption{Evolution of GRB 190114C optical spectral index (not corrected for host galaxy extinction) with the reverse-forward shock models used to fit the optical light curves: in red, the forward shock peaks before observations; in black, the forward shock peaks during observations. T$_0$ corresponds to BAT trigger time.}
\label{fig:LC_GRB190114C_ph}
\end{figure}

Even though both models are compatible with the spectral evolution of the optical index $\beta_{\rm opt} ^{\, \star}$ (see Figure \ref{fig:LC_GRB190114C_ph}), the model with the forward shock peak during MASTER/RINGO3 observations is preferred by early and late-time observations over an early-time forward shock peak (see Table \ref{tab:all_alphas} and Figure \ref{fig:LC_GRB190114C_RINGO3_RSFS_2}). Photoionization of dust could also cause similar color evolution ---  with a red-to-blue shift --- during the very early stages of the GRB and mainly during the prompt phase (e.g., \citealt{2003ApJ...585..775P, 2014MNRAS.440.1810M, 2018ApJS..234...26L}). However, GRB 190114C blue-to-red color change favors the interpretation of the passage of an additional spectral component through the optical band: the transition from reverse shock dominated outflow to forward shock emission (e.g., see GRB 061126; \citealt{2008ApJ...672..449P}, GRB 080319; \citealt{2008Natur.455..183R} and GRB 130427A; \citealt{2014Sci...343...38V}). GRB 061126 from Table \ref{tab:grbs_steep_to_flat_LC} is also identified among the 70 GRBs of \cite{2018ApJS..234...26L} classification of color trends as a reverse to forward shock transition. Additionally, the reverse-forward shock scenario is supported by radio data \citep{2019ApJ...878L..26L}.

\subsection{The Standard Model for a Normal Spherical Decay}

\subsubsection{Evidence of a Jet Break in the X-rays?}  \label{sec:LC_xrays}

\begin{figure}[ht!]
\plotone{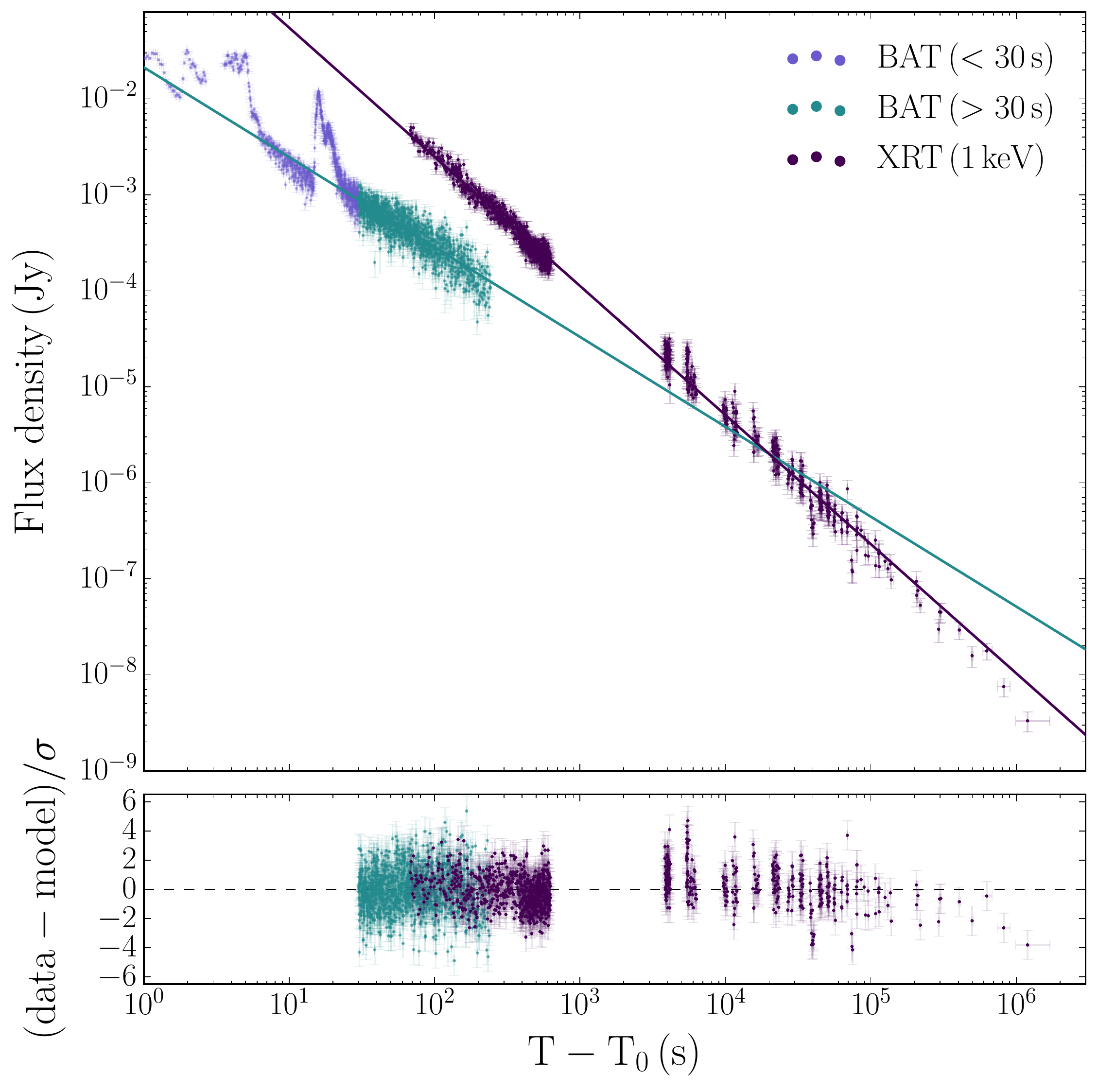}  
\caption{GRB 190114C BAT/XRT emission \citep{2009MNRAS.397.1177E} modeled in terms of power-laws. The bottom panel corresponds to the residuals of the fit and T$_0$ to BAT trigger time.}
\label{fig:LC_GRB190114C_xrt_BAT}
\end{figure}

After the main $\gamma$-ray prompt bulk emission $\gtrsim 30\,$s post-burst, BAT light curve presents a tail of extended emission that we model with a simple power-law until $\sim 240\,$s. This model yields $\alpha_{\gamma} = 0.936 \pm 0.015$ and $\chi ^2 /{\rm d.o.f.} = 2524/1112$ (see Figure \ref{fig:LC_GRB190114C_xrt_BAT}). We notice that a broken power-law model does not increase the significance of the fit.

GRB 190114C X-rays light curve has no shallow phase (see \citealt{2019arXiv191004097Y} for other GeV/TeV events) and decays as $\alpha_{\rm x}= 1.345 \pm 0.004$ through all Swift XRT observations (see Figure \ref{fig:LC_GRB190114C_xrt_BAT}; $\chi ^2 /{\rm d.o.f.} = 1608/1052$), which is similar to the expected $\alpha_{\rm x} \sim 1.2$ decay for the normal spherical stage \citep{2006ApJ...642..389N,2006ApJ...642..354Z}. However, Figure \ref{fig:LC_GRB190114C_xrt_BAT} late-time residuals show signs of a possible break as the XRT light curve model tends to overestimate the flux; the last two observation bins lay $2.6\sigma$ and $3.8\sigma$ away from the chosen model. To account for a possible change of the slope steepness during the late-time afterglow, we fit a broken power-law model which yields $\alpha_{\rm x1} = 1.321 \pm 0.005$, $\alpha_{\rm x2} = 1.49 \pm 0.02$ and a break time at $(1.8 \pm 0.3) \times 10^4\,$s, with $\chi ^2 /{\rm d.o.f.} = 1530/1051$. This means a change of $\Delta \alpha_{\rm x} =0.17 \pm 0.04$ in the temporal decay rate that does not have any spectral break associated; we exclude the passage of a break frequency. For GRB 090102 X-ray afterglow (see Table \ref{tab:grbs_steep_to_flat_LC}), \cite{2010MNRAS.405.2372G} finds a similar temporal break from $\alpha_1=1.29 \pm 0.03$ to $\alpha_2=1.48 \pm 0.10$ at a comparable time $1.9 ^{+1.5} _{-0.8} \times 10^{4} \,$s without any spectral change. Consequently, we explore the possibility of a jet break. From \citealt{1999ApJ...519L..17S} formulation, the jet opening angle is

\begin{equation}
\theta_{\rm j}  \approx 0.0297 \, \Big(\frac{t_{\rm j}}{{\rm 1 \, hr}} \Big)^{3/8}  \, \Big( \frac{E_{\rm iso}}{2.4 \times 10^{53} \, {\rm ergs}} \Big)^{-1/8}   \, \Big( \frac{1+z}{1.4245} \Big)^{-3/8}
\end{equation} for an ISM-like environment and assuming typical values of circumburst density $n=1$ cm$^{-3}$ and radiative efficiency $\eta=0.2$. Taking into account that the jet opening angle distribution of long GRBs peaks around $5.9 \degree$ \citep{2016ApJ...818...18G}, E$_{\rm iso} = (2.4 \pm 0.5) \times 10^{53}\,$erg \citep{2019GCN.23737....1F} and $z=0.4245 \pm 0.0005$ \citep{2019GCN.23708....1C}, the jet break should be visible at $t_{\rm j} \sim 10^{5} \, $s. A jet break at $(1.8 \pm 0.3) \times 10^{4} \, $s --- implying $\theta_{\rm j} \sim 3.1 \degree$ --- is possible and given the scarcity of GCNs observations around the break time, we cannot rule it out.

\subsubsection{The Optical and X-rays Afterglow}

For pure forward shock emission in fireball model conditions, one would expect that if the optical and the X-rays share the same spectral regime, the emission will decay at the same rate. Taking into account that $\alpha_{\rm opt, f}=0.707 \pm 0.010$, $\alpha_{\rm 0.3-10 keV}= 1.345 \pm 0.004$ and $\alpha_{\rm 15-350 keV}=0.936 \pm 0.015$, we find a difference of $\Delta \alpha _{\rm f} =\alpha_{\rm x} -\alpha_{\rm opt, f} = 0.638 \pm 0.011$ between the 0.3-10 keV/optical decay rates and $\Delta \alpha _{\rm f} = 0.23 \pm 0.02$ for the 15-350 keV/optical emission, which implies that there is at least one break frequency in between the X-rays and the optical. This interpretation is also supported by the need of a spectral break between these two bands that changes the slope by $\Delta \beta=0.50 \pm 0.03$ (see Section \ref{sec:SEDs_broadband}).

For ISM medium, slow cooling regime and with the cooling frequency in between the optical and the X-rays bands, an electron index of $p \sim 1.95$ (see Section \ref{sec:reverse}) implies spectral indexes of $\beta_{\rm opt, CR} \sim 0.48$ and $\beta_{\rm x, CR} \sim 0.98$, which are in agreement with $\beta_{\rm opt}=0.43 \pm 0.02$, $\beta_{\rm x}=0.93 \pm 0.02$ derived from the broadband SED modeling (see Section \ref{sec:SEDs_broadband}). The evolution of E$_{\rm  break}$ for the last three SEDs is also consistent with the passage of the cooling frequency $\nu_{\rm c} \propto t^{-\alpha_{\rm c}}$ with $\alpha_{\rm c} \sim 0.5$.

A difference of $\Delta \alpha_{\rm f} = \pm 0.25$ is expected if the cooling frequency lies in between the X-rays/optical bands. Taking into account that $\alpha_{\rm opt, f}=0.707 \pm 0.010$, $\alpha_{\rm 15-350 keV}=0.936 \pm 0.015$ and $\alpha_{\rm 80 keV-8 MeV} = 0.99 \pm 0.05$ \citep{2019GCN.23714....1M}, we find that the 15-350 keV/optical emission $\Delta \alpha _{\rm f} = 0.23 \pm 0.02$ and the 80 keV-8 MeV/optical emission $\Delta \alpha _{\rm f} =0.28 \pm 0.05$ are consistent with $\Delta \alpha_{\rm f} = \pm 0.25$. However, this relation does not hold for the 0.3-10 keV/optical emission with $\Delta \alpha _{\rm f} = 0.638 \pm 0.011$. Furthermore, the steepness of the X-rays light curve $\alpha_{\rm x, f}=1.345 \pm 0.004$ implies a softer $\beta_{\rm x, CR} \sim 1.23$, $p_{\rm x} \sim 2.46$, which does not agree with either the observed spectral indexes or the preferred model for the optical emission. Out of 68 GRBs of \cite{2013A&A...557A..12Z} sample, only $19\%$ of GRBs follow $\Delta \alpha_{\rm f} = 0, \pm 0.25$ for all XRT X-rays/optical light curve segments. GRB 190114C belongs to the $41\%$ of the GRB population that no light curve segments $\Delta \alpha_{\rm f}$ satisfy the fireball model conditions for forward shock emission. Additionally, out of 6 GRBs of \cite{2014ApJ...785...84J} sample with reverse-forward shock signatures, only GRB 090424 fulfils $\Delta \alpha_{\rm f} = 0, \pm 0.25$.

An alternative to reconcile the optical with the soft X-rays emission is to assume that they belong to two spatially or physically different processes. Supporting the scenario of complex jet structure or additional emission components, we have chromatic breaks that cannot be explained either by a break frequency crossing the band or an external density change \citep{2011MNRAS.412..561O}. For example, a two component-jet would produce two forward shocks that would respectively be responsible for the optical and the X-rays emission at late times (GRB 050802; \citealt{2007MNRAS.380..270O}, GRB 080319; \citealt{2008Natur.455..183R}).

\section{Discussion} \label{sec:discussion}

\subsection{Strength of the Magnetic Fields in the Outflow} \label{sec:strength_Bfields}

The reverse shock dynamics have mostly been studied for two regimes \citep{2000ApJ...545..807K}: thick and thin shell. For thick shell regime, the initial Lorentz factor $\Gamma$ is bigger than critical value $\Gamma_{\rm c}$ ($\Gamma > \Gamma_{\rm c}$) and the reverse shock becomes relativistic in the unshocked material rest-frame such that it effectively decelerates the shell. For thin shell regime ($\Gamma \lesssim  \Gamma_{\rm c}$), the reverse shock is sub-relativistic and cannot effectively decelerate the shell. From \cite{2008ApJ...687..443G}, the critical value is 

\begin{equation}
\Gamma_{\rm c} = 258 \bigg(\frac{1+z}{1.4245}\bigg)^{3/8}     \bigg(\frac{T}{30\,{\rm  s}}\bigg)^{-3/8}     \bigg(\frac{E_{\rm iso}}{2.4 \times 10^{53} \, {\rm ergs}}\bigg)^{1/8}
\end{equation}for redshift $0.4245 \pm 0.0005$ \citep{2019GCN.23708....1C}, E$_{\rm iso} = (2.4 \pm 0.5) \times 10^{53}\,$erg \citep{2019GCN.23737....1F}, prompt bulk emission duration T$\,=30\,$s and assuming $n=1\,$cm$^{-3}$.

Our interpretation for GRB 190114C optical afterglow is that the reverse shock peaks at the start or before MASTER observations $\leq {\rm T}_0 + 30\,$s; the early-time observations from MASTER/RINGO3 and late-time GCNs are consistent with the reverse-forward shock model of Figure \ref{fig:LC_GRB190114C_RINGO3_RSFS_2} right, the detection of sub-TeV emission at T$_0 + 50\,$s also supports an early afterglow peak as it is thought to arise from external shocks \citep{2019ATel12390....1M, 2019ApJ...880L..27D} and \cite{2019arXiv190910605A} suggest that the $ \gtrsim {\rm T} _0 + 10\,$s emission has already afterglow contribution. Because the optical afterglow is fading straight after the $\gamma$-ray prompt emission, GRB 190114C should be either in a thick or intermediate regime, $\Gamma \gtrsim \Gamma_c$. For $\Gamma \gg \Gamma_c$, the reverse shock emission should initially decay as $\alpha_{\rm r} \sim 3$ because of the quick energy transfer by a rarefaction wave \citep{2000ApJ...542..819K, 2007ApJ...655..973K}, which is not in agreement with the observations. Consequently, $\Gamma$ should be close the critical value $\Gamma_{\rm c}$, $\Gamma \sim \Gamma_c$; the reverse shock is marginally relativistic at the shock crossing time and the thin shell model is valid.

In order to quantify the strength of the magnetic field in the reverse shock region, \cite{2003ApJ...595..950Z} introduce the magnetic energy ratio $R_{\rm B}$; this parameter is derived assuming different magnetic equipartition parameters for forward $\epsilon_{\rm B, f}$ and reverse shock $\epsilon_{\rm B, r}$ (the fireball ejecta might be endowed with primordial magnetic fields), no or moderate fireball magnetization (the magnetic fields do not affect the fireball dynamics), same electron equipartition parameter $\epsilon_{\rm e}$ and electron index $p$ for both shock regions, thin shell regime and the spectral configuration $\nu_{\rm m,r} < \nu_{\rm m,f}  <  \nu_{\rm c,r} \leq \nu_{\rm c,f}$ at the shock crossing time. Additionally, we assume that the forward shock peaks during RINGO3 observations at $t_{\rm peak, f} \sim 900 \,$s --- masked by reverse shock emission that decays as $\alpha_{\rm opt, r} = 1.711 \pm 0.012$ --- and that the reverse and forward shock emission are comparable at that time. Therefore, \cite{2008ApJ...687..443G} derive

\begin{equation}
R_{\rm B} \equiv \frac{ \epsilon_{\rm B, r}}{  \epsilon_{\rm B, f}}  \sim \Big[R_{\rm t} ^3 \Gamma {^{(4\alpha_{\rm r}-7)}} \Big] ^{2 /(2 \alpha_{\rm r} + 1)},
\end{equation} where $R_{\rm t}$ is the ratio between forward and reverse shock peak times $R_{\rm t} \equiv  t_{\rm peak, f}/t_{\rm peak, r}$. Assuming $\Gamma \sim \Gamma_c$ and $t_{\rm peak, r} \sim 30\,$s, we estimate that the magnetic energy density in the reverse shock region is higher than in the forward shock by a factor of $R_{\rm B} \sim 70$; the reverse shock emission could have globally ordered magnetic fields advected from the central engine.

Broadband afterglow modeling usually shows levels of $\epsilon_{\rm B, f} \sim 10^{-5} - 10^{-1}$ for the forward shock magnetic equipartition parameter \citep{2002ApJ...571..779P}. For GRB 190114C, $\epsilon_{\rm B, f} \sim 10^{-5}-10^{-4}$ \citep{2019ApJ...884..117W, 2019arXiv190910605A,2019ApJ...879L..26F}; so this GRB is likely weakly magnetized at the deceleration radius. Consequently, magnetic fields are dynamically subdominant and bright reverse shock emission is expected \citep{2003ApJ...595..950Z, 2004A&A...424..477F, 2005ApJ...628..315Z}. If $\epsilon_{\rm B, f} \sim 0.1$ --- as discussed in \citealt{2019ApJ...880L..27D} --- $\sigma_{\rm B}$ would be order of unity and our model assumption (i.e magnetic fields
do not affect the dynamics of the outflow) becomes invalid. Although reconnections might be able to produce the prompt and early afterglow emission
in the high magnetization regime (e.g., \citealt{2001AandA...369..694S, 2003astro.ph.12347L, 2011ApJ...726...90Z}), our forward-reverse shock model (purely hydrodynamics model) can describe the early afterglow well and we assume $\epsilon_{\rm B, f} \sim 10^{-5}-10^{-4}$ as our fiducial value.

\subsection{Maximum Reverse Shock Synchrotron-Self-Compton Energy}

The maximum synchrotron energy that can be produced by shock-accelerated electrons is about $\nu_{\rm max}^\prime \sim m_{\rm e} c^2/\alpha_{\rm FS} \sim 100$ MeV in the shock comoving frame where $\alpha_{\rm FS}$ is the fine-structure constant. For the observer, this limit is boosted by the bulk Lorentz factor as $\nu_{\rm max} \sim 100 \Gamma $ MeV $\sim 20 (\Gamma / 200)$ GeV. Since the bulk Lorentz factor is less than a few hundred in the afterglow phase, Synchrotron Self-Compton (SSC) processes are favored to explain the sub-TeV emission \citep{2019ApJ...880L..27D, 2019arXiv190910605A,2019ApJ...883..162F,2019arXiv191014049Z, 2019A&A...626A..12R}.

Considering the longevity of the high-energy emission, the SSC emission is likely to originate from the forward shock region. As we discuss below, the maximal Inverse Compton photon energy also favors the forward shock origin.
 
The typical random Lorentz factor of electrons in the reverse shock region is about $\gamma_{\rm m,r} \sim (\epsilon_{\rm e} /3)(m_{\rm p} /m_{\rm e})\sim 20 (\epsilon_{\rm e}/3\times10^{-2}) $ at the onset of the afterglow, and it cools due to the adiabatic expansion of the shock ejecta as $\gamma_{\rm m,r} \propto t^{-2/7}$ \citep{2000ApJ...545..807K}. Since the typical value is lower by a factor of order $\Gamma$ than that in the forward shock region, it is difficult to produce very high energy emission in the reverse shock region even if a higher-order inverse Compton (IC) component is considered \citep{2007ApJ...655..391K}. If the intermediate photon energy in the higher-order IC scattering (i.e. the photon energy before the scattering in the electron comoving frame) is too high, the Klein-Nishina effect suppresses the higher-order IC scattering. Since the intermediate photon energy can be as high as $\sim 100$ keV  $(\ll m_{\rm e} c^2)$ and still be in the Thomson limit, the maximum IC energy is at most 100 $\gamma_{\rm m,r}\Gamma$ keV $\sim 3 (\gamma_{\rm m,r}/100) (\Gamma/300) $ GeV. Basically, the same limit can be obtained by considering that electrons with random Lorentz factor $\gamma_{\rm e}$ should be sufficiently energetic $\Gamma \gamma_{\rm e} m_{\rm e} c^2 \gg h\nu_{\rm IC}$ to upscatter a low-energy photon to a high-energy $h\nu_{\rm IC}$.

\subsection{Structure of the Magnetic Fields in the Outflow}

Whilst the magnetization degree determines the strength of the magnetic field, GRB linear polarimetry directly informs of the degree of ordered magnetic fields in the emitting region (e.g., length scales and geometry).

Theoretically, synchrotron emission can be up to $70\%$ polarized \citep{1979rpa..book.....R}, but this can be further reduced due to: inhomogeneous magnetic fields (e.g., highly tangled magnetic fields, patches of locally ordered magnetic fields), a toroidal magnetic field viewed with a line-of-sight almost along the jet axis, the combination of several emission components endowed with ordered magnetic fields but with different polarization components (e.g., internal-external shocks) or the combination of reverse-forward shock emission. Additionally, if the reverse shock is propagating in a clumpy medium, polarization levels could be also reduced \citep{2017ApJ...845L...3D}. If the emission region contains several independent patches of locally ordered magnetic fields, the degree and direction of polarization should depend on time as the process is stochastic.

In Section \ref{sec:after_model}, we have discussed that the steep-to-flat behavior of GRB 190114C optical light curve is most likely due to a reverse-forward shock interplay. If the reverse shock emission is highly polarized, the degree of polarization should decline steadily as the unpolarized forward shock emerges (GRB 120308A; \citealt{2013Natur.504..119M}). In GRB 190114C, the reverse shock dominates the afterglow emission from $52\,$s to $109\,$s post-burst and the polarization degree drops abruptly from $7.7 \pm 1.1\%$ to $2.0 ^{+2.6} _{-1.5} \%$. From $200\,$s to $\sim 2000\,$s post-burst, the fraction of reverse to forward shock flux density declines from $\sim 0.96$ to $\sim 0.31$ and we detect $2- 4\%$ constant polarization degree in all three RINGO3 bands throughout this period. This contrasts with the higher value ${\rm P}=10 \pm 1 \%$ measured during the early light curve of GRB 090102 \citep{2009Natur.462..767S, 2010MNRAS.405.2372G}, which shows a similar light curve behavior of steep-to-flat decay typical of a combination of reverse and forward shock emission. At the polarization observing time, the modeling of GRB 090102 afterglow ($\alpha_{\rm r}=1.987 \pm 0.012$, $t_{\rm peak, f} = 205 \pm 38\,$s) indicates that the proportion of reverse to forward shock emission was $\sim 0.58$, implying that the intrinsic polarization of the reverse shock emission is higher than the observed (i.e. the ejecta contains large-scale ordered magnetic fields). GRB 190114C polarization properties are also markedly different to those of GRB 120308A in which the observed reverse shock emission is dominant and highly polarized ($28 \pm 4\%$) at early times, decreasing to $16 ^{+5} _{-4} \%$ as the forward shock contribution increases with time.

In short, the polarization of the optical emission in GRB 190114C is unusually low despite the clear presence of a reverse shock. We suggest the initial $7.7 \pm 1.1\%$ and sudden drop to $2.0 ^{+2.6} _{-1.5} \%$ may be due to a small contribution from optically polarized prompt photons (as for GRB 160625B; \citealt{2017Natur.547..425T}) but therefore the dominant polarization degree of the afterglow is between $2-4\%$ throughout. We next discuss possible scenarios to explain this low and constant $2-4\%$ degree.

\subsubsection{Dust-induced Polarization: Low Intrinsic Polarization in the Emitting Region}

GRB 190114C is a highly extincted burst, which complicates polarization measurements intrinsic to the afterglow. Because of the preferred alignment of dust grains, dust in the line-of-sight can induce non-negligible degrees of polarization that vectorially add to the intrinsic afterglow polarization; late-time polarimetric studies of GRB afterglows show few percents of polarization (e.g., \citealt{1999A&A...348L...1C, 2004ASPC..312..169C, 2004AIPC..727..269G,2012MNRAS.426....2W}). For GRB 190114C line-of-sight, the polarization of CD-27 1309 star P$_{\rm \lbrace BV, R, I \rbrace} = 0.3 \pm 0.1\%, 0.1 \pm 0.1\%, 0.3 \pm 0.1\%$ gives an estimation of the polarization induced by Galactic dust. For the host galaxy, we estimate the dust-induced polarization degree with the Serkowski empirical relation \citep{1975ApJ...196..261S, 1992ApJ...386..562W} 

\begin{equation}
{\rm P} = {\rm P}_{\rm max} \exp \Big[-K \ln^2  \Big( \frac{\lambda_{\rm max}}{\lambda}  \Big)\Big],
\end{equation}where $\lambda_{\rm max}({\rm \mu m})=R_{\rm V}/5.5$, $K = 0.01 \pm 0.05 + (1.66 \pm 0.09)\lambda_{\rm max}$ and P$_{\rm max} \lesssim 9 \, $E$_{\rm B-V}$. We introduce the redshifted-host effect $\lambda_{\rm max} \longrightarrow (1+z) \lambda_{\rm max, \, HG}$ \citep{2004AandA...420..899K, 2012MNRAS.426....2W} and we assume MW extinction profile with E$_{\rm B-V, MW} =0.0124 \pm 0.0005$ \citep{1998ApJ...500..525S} and SMC profile for the host galaxy with E$_{\rm B-V, HG}=0.51 \pm 0.04$. Taking into account the shape of RINGO3 bandpasses, we find that the maximum polarization degree induced by the host galaxy dust is P$_{\rm \lbrace BV, R, I \rbrace}  \lesssim 3.9\%, 4.5\%, 4.5\%$, compatible with the constant $2-4\%$ polarization degree of the GRB detected since $109\, $s post-burst.

Depending on the relative position of the polarization vectors (the alignment of dust grains to the intrinsic polarization vector of the ejecta), dust could either polarize or depolarize the outflow. If dust was depolarizing the intrinsic polarization, this would mean a gradual rotation of the angle as the percentage of polarized reverse shock photons decrease. The constant angle and polarization degree favors the interpretation that the $\sim 2-4\%$ ordered component is compatible with dust-induced levels (see Figure \ref{fig:P_evolution}); i.e. the intrinsic polarization at that time is very low or negligible.

\subsubsection{Distortion of the Large-Scale Magnetic Fields}

Although the early afterglow modeling implies that the ejecta from the central engine is highly magnetized for this event, the polarization degree of the reverse shock emission is very low and the $2-4 \%$ polarization signal is likely to be induced by dust. This is in contrast to the high polarization signals observed in other GRB reverse shock emission (GRB 090102; \citealt{2009Natur.462..767S}, GRB 101112A; \citealt{2017ApJ...843..143S}, GRB 110205A; \citealt{2017ApJ...843..143S}, GRB 120308A; \citealt{2013Natur.504..119M}).

One possibility is that the low degree of polarization arises from other emission mechanisms in addition to synchrotron emission. Since the optical depth of the ejecta is expected to be well below unity at the onset of afterglow, most synchrotron photons from the reverse shock are not affected by IC scattering processes (the cooling of electrons is also not affected if the Compton y-parameter is small). The polarization degree of the synchrotron emission does not change even if the IC scattering is taken into account. However, the polarization degree is expected to be reduced for the photons upscattered by random electrons (i.e. SSC photons; \citealt{2017ChPhC..41d5101L}). We now consider whether this can explain the observed low polarization degree of the reverse shock emission. 

If the typical frequency of the forward shock emission is in the optical band $\nu_{\rm m,f}\sim 5 \times 10^{14}\,$Hz at $t \sim 900\,$s as our afterglow modeling suggests (the right panel of Figure \ref{fig:LC_GRB190114C_RINGO3_RSFS_2}), it should be about $\nu_{\rm m,f} \sim 8\times 10^{16}\,$Hz at the onset of afterglow ($t_{\rm d} \sim 30\,$s). Since the typical frequency of the reverse shock emission is lower by a factor of $\sim \Gamma^2$ (this factor weakly depends on the magnetization parameter $R_{\rm B}$, but the inclusion of a correction factor does not change our conclusion; see \cite{2013ApJ...772..101H} for more details), it is about $\nu_{\rm m,r} \sim 10^{12}\,$Hz at that time for $\Gamma \sim \Gamma_{\rm c}=260$. Assuming random Lorentz factor of electrons in the reverse shock region $\gamma_{\rm m,r}  \sim 20 (\epsilon_{\rm e} / 3 \times 10^{-2})$, the typical frequency of the 1st SSC emission is in the optical band $\nu^{\rm IC}_{\rm m,r} \sim \gamma_{\rm m,r}^2 \nu_{\rm m,r} \sim 5 \times 10^{14}\, $Hz.

The optical depth of the ejecta at the onset of afterglow is given by $\tau = \sigma_{\rm T} N_{\rm e} / 4 \pi R_{\rm d} ^2  \sim (\sigma_{\rm T} / 3) \Gamma n R_{\rm d} \sim  7 \times 10^{-6} n$ where $\sigma_{\rm T}$ is the Thomson cross section, $N_{\rm e}$ is the number of electrons in the ejecta, $R_{\rm d} \sim 2 c  \Gamma^2 t_{\rm d} \sim 10^{17} \, $cm is the deceleration radius, and we have used the fact that the mass of the ejecta is larger by a factor of $\Gamma$ than that of the ambient material swept by the shell at the deceleration time. The spectral peak power of the 1st SSC emission is roughly given by $F_{\rm max}^{\rm IC} \sim \tau F_{\rm max,r}$ where $F_{\rm max,r}$ is the spectral peak power of the reverse shock synchrotron emission (e.g., \citealt{2007ApJ...655..391K}). The ratio of the contributions from the 1st SSC and the synchrotron emission to the optical band is about $\tau F_{\rm max,r}/(F_{\rm max,r} (\nu_{\rm opt}/\nu_{\rm m,r})^{-(p-1)/2}) \sim \tau (\nu_{\rm opt}/\nu_{\rm m,r})^{1/2} \sim 7 \times 10^{-6} n$ at the onset of the afterglow. Since the synchrotron emission dominates the optical band, the IC process does not explain the low polarization degree.

Consequently, we suggest that GRB 190114C large-scale ordered magnetic fields could have been largely distorted on timescales previous to reverse shock emission (see also GRB 160625B; \citealt{2017Natur.547..425T}). We speculate that the detection of bright prompt and afterglow emission from TeV to radio wavelengths in GRB 190114C, coupled with the low degree of observed optical polarization, may be explained by the catastrophic/efficient dissipation of magnetic energy from and consequent destruction of order in primordial magnetic fields in the flow; e.g., via turbulence and reconnection at prompt emission timescales (ICMART; \citealt{2011ApJ...726...90Z,2015ApJ...805..163D,2016ApJ...821L..12D, 2016MNRAS.456.1739B}). For GRB 190114C, reconnection could be a mechanism for the production of the high-energy Fermi-LAT photons that exceed the maximum synchrotron energy (another possibility is SSC; \citealt{2019arXiv190910605A}). If the $7.7 \pm 1.1\%$ detection at $52\, $s post-burst is interpreted as due to a residual contribution from polarized prompt photons (as in GRB 160625B; \citealt{2017Natur.547..425T}), this would further support the existence of ordered magnetic fields close to prompt emission timescales and their consequent destruction for reverse shock emission.

The sample of high-quality early time polarimetric observations of GRB afterglows remains small ($<10$) and for prompt emission, smaller still (2). Future high quality early time polarimetric observations at optical and other wavelengths are vital to determine the intrinsic properties of GRB magnetic fields and their role in GRB radiation emission mechanisms.

\section{Conclusions} \label{sec:conclusions}

The early-time optical observations of GRB 190114C afterglow yields an important constraint on the shock evolution and the interplay between reverse and forward shock emission. The steep-to-flat light curve transition favors the presence of reverse shock emission with the forward shock peaking during RINGO3 observations.

The forward-reverse shock modeling suggests that the microscopic parameter $\epsilon_{\rm B}$ is higher by a factor of  $\sim 70$ in the reverse shock than in the forward shock region. It indicates that the fireball ejecta is endowed with the primordial magnetic fields from the central engine. Since we have successfully modeled the early afterglow in the forward-reverse shock framework, the outflow is likely to be baryonic rather than Poynting-flux-dominated at the deceleration radius.

GRB 190114C polarization degree undergoes a sharp drop from $7.7 \pm 1.1\%$ to $2.0 ^{+2.6} _{-1.5} \%$ during $52-109\, $s post-burst not consistent with pure reverse shock emission; we suggest a contribution from prompt photons. Later on, multi-band polarimetry also shows constant ${\rm P} = 2 - 4 \%$ polarization degree during the reverse-forward shock interplay consistent with dust-induced levels from the highly extincted host galaxy. The low intrinsic polarization signal is in contrast to ${\rm P} > 10\%$ measured previously for the events which show a signature of reverse shock emission (i.e. steep rise or decay). Forward shock SSC emission is favored for the origin of the long-lasting sub-TeV emission (we have shown that reverse shock SSC is not energetic enough to produce the sub-TeV emission). We have also tested whether reverse shock SSC emission can explain the low optical polarization degree --- the polarization degree of the photons upscattered by random electrons would be lower than that of the synchrotron photons. Since we show that the 1st SSC component in the optical band is masked by the synchrotron component, the IC process does not explain the low polarization degree. Instead, the unexpectedly low intrinsic polarization degree in GRB 190114C can be explained if large-scale jet magnetic fields are distorted on timescales prior to reverse shock emission.

A larger, statistical sample of early-time polarization measurements with multi-wavelength information is required to understand timescales and mechanisms that cause distortion of the large-scale ordered magnetic fields and ultimately constrain jet models.

\acknowledgments

We thank the anonymous referee for a constructive report that improved the quality of our paper. The Liverpool Telescope is operated on the island of La Palma by Liverpool John Moores University in the Spanish Observatorio del Roque de los Muchachos of the Instituto de Astrofisica de Canarias with financial support from the UK Science and Technology Facilities Council. MASTER is supported in parts by Development program of Lomonosov Moscow State University (equipment), by RFBR grant 19-29-11011. This work made use of data supplied by the UK Swift Science Data Centre at the University of Leicester. The research leading to these results has received funding from the European Union's Horizon 2020 Programme under the AHEAD project (grant agreement 654215). N.J. and C.G.M. acknowledge financial support from Mr Jim Sherwin and Mrs Hiroko Sherwin. C.G.M. acknowledges support from the Science and Technology Facilities Council and the UK Research and Innovation (ST/N001265/1). C.G. acknowledges support for this work provided by Universit\`a di Ferrara through grant FIR 2018 \textquote{A Broad-band study of Cosmic Gamma-Ray Burst Prompt and Afterglow Emission}. A.G. acknowledges the financial support from the Slovenian Research Agency (grants P1-0031, I0-0033, and J1-8136) and networking support by the COST Actions CA16104 GWverse and CA16214 PHAROS. D.A.H.B. acknowledges research support from the South African National Research Foundation. N.B. was supported by RFBR BRICS grant 17-52-80133 and by the Russian Federation Ministry of Science and High Education (agreement N$^{\rm \circ}$ 075-15-2019-1631).

\software{Matplotlib \citep{2007CSE.....9...90H}, SciPy \citep{2019arXiv190710121V}, PyFITS \citep{1999ASPC..172..483B}, Astropy Photutils \citep{2016ascl.soft09011B}, Astroalign \citep{2019arXiv190902946B}, XSPEC and PyXSPEC (v. 12.9.1; \citealt{1999ascl.soft10005A}), HEAsoft (v6.22.1; \citealt{1995ASPC...77..367B}).}

\bibliographystyle{aasjournal}
\bibliography{sample63}{}

\end{document}